%% file: ANA-EXOT-2018-61-PAPER.tex
\newcommand*{\ATLASLATEXPATH}{}
\documentclass[cernpreprint, atlasdraft=false, texlive=2016, USenglish]{\ATLASLATEXPATH atlasdoc}
 
\usepackage{\ATLASLATEXPATH atlaspackage}
\usepackage{\ATLASLATEXPATH atlasbiblatex}
 
\usepackage{\ATLASLATEXPATH atlasphysics}
\usepackage[]{caption}
\usepackage[]{threeparttable}
 
\graphicspath{{logos/}{figures/}}
 
\usepackage{comment}
\usepackage{multirow}
\usepackage{xcolor,colortbl}
\definecolor{Gray}{gray}{0.85}
\definecolor{LightCyan}{rgb}{.75,1,1}
\newcolumntype{a}{>{\columncolor{Gray}}c}
\newcolumntype{b}{>{\columncolor{LightCyan}}c}
 
\AtlasTitle{Search for long-lived neutral particles produced in $pp$ collisions at ${\sqrt{\mathrm s}} = 13~\TeV$ decaying into displaced hadronic jets in the ATLAS inner detector and muon spectrometer}

\AtlasJournalRef{\PRD 101 (2020) 052013}
\AtlasDOI{10.1103/PhysRevD.101.052013}
\AtlasAbstract{
A search is presented for pair-production of long-lived neutral particles using 33 $\mathrm{fb}^{-1}$ of ${\sqrt{\mathrm s}} = 13~\TeV$ proton--proton collision data, collected during 2016 by the ATLAS detector at the LHC.
This search focuses on a topology in which one long-lived particle decays in the ATLAS inner detector and the other decays in the muon spectrometer.
Special techniques are employed to reconstruct the displaced tracks and vertices in the inner detector and in the muon spectrometer.
One event is observed that passes the full event selection, which is consistent with the estimated background.
Limits are placed on scalar boson propagators with masses from $125 \, \GeV$ to $1000 \, \GeV$ decaying into pairs of long-lived hidden-sector scalars with masses from $8 \, \GeV$ to $400 \, \GeV$.
The limits placed on several low-mass scalars extend previous exclusion limits in the range of proper lifetimes $c \tau$ from 5~cm to 1~m.
}
 
\author{The ATLAS Collaboration}
 
\PreprintIdNumber{CERN-EP-2019-240}
 
\AtlasRefCode{EXOT-2018-61}

\AtlasCoverSupportingNote{Support Documentation}{https://cds.cern.ch/record/2311382}
 
\AtlasCoverCommentsDeadline{31 October 2019}
 
\AtlasCoverAnalysisTeam{Benjamin Brau, Jackson Burzynski, Margaret Lutz}
 
\AtlasCoverEdBoardMember{Matthew~Gignac, Laura~Jeanty~(chair), Heather~Russell}

\AtlasCoverEgroupEditors{atlas-EXOT-2018-61-editors@cern.ch}
 
\AtlasCoverEgroupEdBoard{atlas-EXOT-2018-61-editorial-board@cern.ch}

\hypersetup{pdftitle={ATLAS document},pdfauthor={The ATLAS Collaboration}}
 
\begin{document}
 
\maketitle
 
\tableofcontents

\section{Introduction}
\label{sec:intro}

Long-lived particles (LLPs) are predicted by many extensions of the Standard Model (SM), including various supersymmetric (SUSY)~\cite{Giudice:1998bp,Barbier:2004ez,Fan:2012jf,Csaki:2015fea}, hidden sector (HS)~\cite{EchoesHV,HV_DV,HVPhenoHadronColl} and neutral naturalness~\cite{Chacko:2005pe,Cai:2008au,Chacko:2015fbc} models that address the hierarchy problem.
Decays of LLPs may go unnoticed in searches at collider experiments that are designed to identify promptly decaying particles.
Searches for LLPs provide a promising avenue for the discovery of physics beyond the Standard Model (BSM).
The search presented in this paper is sensitive to neutral LLPs that are pair-produced, with one LLP decaying in the ATLAS inner tracking detector (ID) and the other in the muon spectrometer (MS).
This particular event topology provides sensitivity to LLPs with proper lifetimes ($c\tau$) ranging from a few centimeters to several meters.
 
\begin{figure}[!htb]
\includegraphics[width=0.5\textwidth]{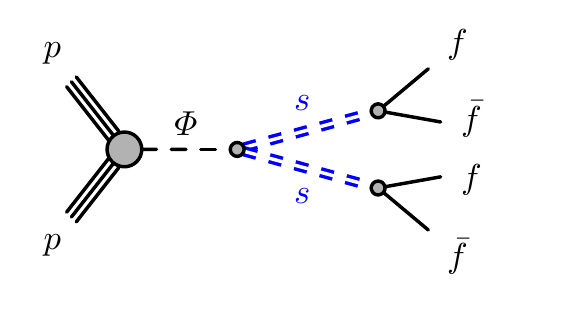}
\centering
\caption{Diagram for a Higgs boson or heavy scalar $\Phi$ decaying into displaced hadronic jets via a hidden sector.}
\label{fig:HSSFeyn}
\end{figure}

In HS models, a set of BSM particles is weakly coupled to the SM via a mediator particle.
These models are intriguing because they can be built in multiple ways and can produce LLPs with little to no fine-tuning~\cite{HV_DV}.
A SM Higgs boson mediator is of particular interest because the current experimental characterization of the Higgs boson allows sizable couplings of the Higgs boson to the BSM sector~\cite{Aad:2019mbh,HIGG-2015-03}.
HS models are also compatible with SUSY~\cite{HV_DV,Strassler:2006qa} and with models of neutral naturalness~\cite{Curtin:2015fna}.
The results of this search are interpreted in the context of a simplified HS model, in which the SM and HS are connected via a heavy mediator $\Phi$, which decays into a pair of long-lived neutral scalar particles $s$ as shown in Figure~\ref{fig:HSSFeyn}.
The $s$-bosons then decay back into SM particles through their mixing with the mediator \cite{Curtin:2013fra}.
The HS model assumes an effective Yukawa coupling between the $s$-boson and the SM particles; therefore each $s$-boson decays primarily into a heavy fermion pair: $b\bar{b}$, $c\bar{c}$, $\tau^{+}\tau^{-}$.
The branching ratio depends on the mass of the $s$-boson ($m_s$), but for $m_s > 25~\GeV$, the branching ratio is approximately 85:5:8.\footnote{
As in recent displaced jet searches for HS models~\cite{CalRatio2016,EXOT-2017-05}, decays of $s$ to bosons and top quarks are not considered in this analysis.
}
If the SM particles are quarks, they hadronize, resulting in jets that may be highly displaced from the interaction point (IP).
The proper lifetime of the $s$-boson is relatively unconstrained aside from the upper limit imposed by Big Bang nucleosynthesis of $c\tau \lesssim 10^8$~m~\cite{Jedamzik:2006xz}.
 
Searches for displaced hadronic decays have been performed by ATLAS, CMS, and LHCb in the first run of the LHC~\cite{EXOT-2011-11,LHCb_7TeV_DV,EXOT-2013-12,SUSY-2014-02,CMS-EXO-12-038,EXOT-2012-28,Aaij:2017mic,Aaij:2016xmb},
as well as in the second run of the LHC by CMS~\cite{CMS-EXO-16-003} and ATLAS~\cite{SUSY-2016-08,EXOT-2017-03,EXOT-2017-05,CalRatio2016}.
Two searches for displaced decays resulting from neutral LLPs in the HS model have been performed using the ATLAS Run 2 dataset.
One analysis searched for pairs of displaced hadronic jets in the calorimeter~\cite{CalRatio2016} (the CR analysis) and the other searched for one or two displaced hadronic jets in the MS~\cite{EXOT-2017-05} (the MS analysis).
For a SM Higgs boson mediator with $m_H = 125$~GeV, decays of neutral scalars with masses between 8 and 55~GeV have been excluded by these two analyses for $c\tau$ between 7~cm and 220~m depending on the LLP mass (assuming a 10\% branching ratio of the Higgs boson into $ss$ pairs).
 
This analysis uses 33.0~$\mathrm{fb}^{-1}$ of $\sqrt{s} = 13~\TeV$ proton--proton ($pp$) collision data collected by the ATLAS detector at the LHC and is an update to the results presented in the 8 TeV ATLAS search~\cite{EXOT-2013-12} for displaced hadronic jets in the ID and MS, and an extension of the MS analysis.
In each event, one reconstructed decay vertex is required in the ID (IDVx) in addition to one in the MS (MSVx).
Requiring the presence of an MSVx suppresses background, which allows looser selection requirements on the reconstructed mass and number of tracks associated with the IDVx than in other IDVx searches such as Ref.~\cite{SUSY-2016-08}.
Additionally, requiring the presence of an IDVx suppresses background relative to the MS analysis, allowing greater sensitivity for LLPs with $c\tau < 1$~m.
This search thus increases the sensitivity to low-mass scalars with shorter proper lifetimes relative to the combined results of the CR and MS analyses published in Ref.~\cite{CalRatio2016}.
 
A signature-driven trigger that is used to collect data selects candidate events for decays of LLPs in the MS~\cite{TRIG-2012-02,EXOT-2017-05}.
Standard ATLAS reconstruction methods are optimized for prompt decays and may fail to reconstruct the decays of long-lived particles.
Thus, specialized tracking and vertex reconstruction algorithms are used that allow for the reconstruction of displaced decays in the ID~\cite{ATL-PHYS-PUB-2017-014,PERF-2011-08,SUSY-2016-08} and the MS~\cite{PERF-2013-01}.
 
The main background source of signal-like vertices in the ID is the interactions between SM particles and the material in the inner detector, which may create multi-track vertices that
are displaced from the IP.
Other sources of background are hadronic jets, vertices reconstructed from fake tracks that are created from multiple unrelated energy deposits, and vertices reconstructed from random track crossings.
The main source of background vertices in the MS is electromagnetic or hadronic showers that are not contained in the calorimeters (\textit{punch-through jets}).
Other sources of background in the MS are multijet events with mismeasured jets and non-collision backgrounds such as machine-induced background~\cite{BRUCE2013825,DAPR-2017-01}, electronic noise and cosmic-ray muons.
A data-driven background estimation is used to determine how many events in the signal region result from background sources.
 
This paper is organized as follows.
The ATLAS detector is described in Section~\ref{sec:detector}.
The data and simulated samples used are outlined in Section~\ref{sec:data}, and the trigger used to collect the data is detailed in Section~\ref{sec:trig}.
The algorithms used for the displaced track and vertex reconstruction are explained in Section~\ref{sec:reco}, followed by a discussion of the selection criteria for the events and vertices in Section~\ref{sec:select}.
The background estimation procedure is described in Section~\ref{sec:background} and the systematic uncertainties are summarized in Section~\ref{sec:syst}.
Finally, the results and conclusion are presented in Sections~\ref{sec:result} and~\ref{sec:conclusion}.
 
\section{ATLAS detector}
\label{sec:detector}
 
The ATLAS
detector~\cite{PERF-2007-01,ATLAS-TDR-19,PIX-2018-001} at
the LHC is a cylindrical multipurpose particle detector with forward--backward symmetry and nearly $4\pi$ solid angle coverage.
The detector is composed of the inner tracking detector, the electromagnetic and hadronic calorimeters, and the muon spectrometer.
 
The ID covers the range $0.03 \text{ m} < r < 1.1 \text{ m}$ and $|z| < 3.5$~m,\footnote{ ATLAS uses a right-handed coordinate system with its origin at the nominal interaction point in the center of the detector and the $z$-axis along the beam pipe. The $x$-axis points from the IP to the center of the LHC ring, and the $y$-axis points upwards. Cylindrical coordinates $(r,\phi)$ are used in the transverse plane, $\phi$ being the azimuthal angle around the $z$-axis. Pseudorapidity is defined in terms of the polar angle $\theta$ as $\eta = - \ln \tan(\theta/2)$. Angular distance is measured in units of $\Delta R \equiv \sqrt{ \left(\Delta \eta \right)^2 + \left(\Delta \phi \right)^2   }$.} and is immersed in a 2~T axial magnetic field from a superconducting solenoid.
Three ID sub-detectors provide precision tracking for charged particles within the pseudorapidity region $|\eta| < 2.5$.
At small radii, a silicon pixel sub-detector provides high-resolution position measurements.
The pixel system consists of four barrel layers positioned at radii of 33.3~mm, 50.5~mm, 88.5~mm, and 122.5~mm, and three forward disks in each endcap.
The pixel detector is surrounded by the silicon microstrip tracker (SCT), which consists of four double layers in the barrel and nine forward disks in each endcap.
The radial position of the first (last) SCT layer is 299~mm (514~mm).
The outermost sub-detector of the ID is the straw-tube transition radiation tracker (TRT), which provides an average of 30 additional two-dimensional points to tracks in the range $|\eta| < 2.0$.
 
The calorimeter system provides coverage over the pseudorapidity range $|\eta| < 4.9$.
It consists of an electromagnetic calorimeter surrounded by a hadronic calorimeter.
Within the region $|\eta| < 3.2$, the electromagnetic calorimeter consists of liquid-argon (LAr) barrel and endcap sampling calorimeters with lead absorbers.
The hadronic calorimeter consists of steel/scintillator tile calorimeters within $|\eta| < 1.7$, and two copper/LAr hadronic endcap calorimeters that cover the region $1.5 < |\eta| < 3.2$.
A forward calorimeter using copper and tungsten absorbers with LAr completes the calorimeter coverage up to $|\eta| = 4.9$.
 
The MS consists of separate trigger and high-precision tracking chambers measuring the deflection of muons in a magnetic field generated by superconducting air-core toroids.
The muon tracking chamber system covers the region $|\eta| < 2.7$ with three layers of monitored drift tubes (MDT), complemented by cathode strip chambers (CSC) in the forward region.
The MDT chambers consist of two multilayers, each of which consists of three or four layers of drift tubes.
Three stations of resistive plate chambers (RPC) and thin gap chambers (TGC) are used for triggering and $\phi$ measurements in the MS barrel and endcaps, respectively.
The muon trigger system covers the range $|\eta| < 2.4$.
 
The ATLAS detector selects events using a two-tiered trigger system~\cite{TRIG-2016-01}.
The first level (L1), which is a hardware-based system, uses coarse data collected from the calorimeters and muon detectors to reduce the event rate from the LHC crossing frequency of 40~MHz to a design value of 100~kHz. The second level, known as the high-level trigger (HLT), is a software-based system that uses information from all of the ATLAS sub-detectors to reduce the rate of recorded events to approximately 1~kHz.
 
\section{Data and simulated events}
\label{sec:data}
 
\subsection{Data events}
 
This analysis uses 33.0~$\mathrm{fb}^{-1}$ of data collected by the ATLAS detector during the 2016 data-taking period using $pp$ collisions at $\sqrt{s} = 13$~\TeV.
The analysis is performed using selected subsets of data which underwent special reconstruction of displaced tracks and vertices.
 
Two sets of data are used.
One set consists of events that pass a signature-driven LLP trigger, referred to here as the Muon RoI Cluster trigger (described in Section~\ref{sec:trig}).
The Muon RoI Cluster trigger is used to collect events for the signal region; these events are also required to contain reconstructed displaced vertices in the MS and the ID, described in Sections~\ref{sec:reco} and~\ref{sec:select}.
Events collected by the Muon RoI Cluster trigger are also used to define a validation region for the background estimation, as discussed in Section~\ref{sec:background}, in which case the event selection is agnostic to the presence of any reconstructed vertices in the MS.
 
The second set of data is used for the background estimation.
Events are selected by a single-muon trigger that requires a muon with transverse momentum $\pt > 26~$\GeV~at the HLT.
To reduce the signal contamination, these events are also required to contain isolated muons.
This requirement is further described in Section~\ref{sec:background}.
 
All of the events that pass the triggers used by the CR analysis~\cite{CalRatio2016}, which select events with displaced hadronic jets decaying in the hadronic calorimeter, are vetoed in this analysis.
This veto is imposed in order to facilitate the combination of the results of this search with the results of the CR and MS analyses (presented in Section~\ref{sec:result}).
 
\subsection{Simulated events}
 
The signal Monte Carlo (MC) samples were generated with a $\Phi$ mediator connecting the SM to a hidden sector in which $\Phi$ decays into pairs of long-lived neutral scalars $ss$.
The neutral scalars $s$ decay back into the SM via their coupling to the mediator, assuming a Yukawa coupling between the $s$-boson and the SM particles.
The $\Phi \rightarrow ss$ signal samples were generated using \textsc{MadGraph5}~\cite{Alwall:2014hca_fix}.
Events were showered with \textsc{Pythia} 8.210~\cite{Sjostrand:2014zea} using the A14 set of tuned parameters (tune)~\cite{ATL-PHYS-PUB-2014-021} and the NNPDF2.3LO set~\cite{Ball:2012cx} of parton distribution functions (PDF).
Various mass points were generated corresponding to different combinations of the mass of the $\Phi$ ($m_{\Phi}$) and $m_s$, with $m_{\Phi} \in [125,1000]$~\GeV~and $m_s \in [8,400]$~\GeV.
The sensitivity of the analysis to models with $s$-boson mass smaller than 8 GeV is limited by the ID vertex reconstruction efficiency and selections to discriminate against vertices from background.
 
The LLP proper lifetime in each sample was tuned so that each mass point had an approximate \textit{mean lab-frame decay length} of 5~m.
The mean lab-frame decay length of 5~m is used so that there are approximately equal numbers of LLP decays in the ID and in the MS, and samples with this lab-frame decay length are used for the determination of the signal versus background selection for the IDVx, as well as the IDVx and overall efficiency studies.
The overall efficiency for a range of proper lifetimes is estimated by reweighting the 5~m samples using an extrapolation method, as noted in  Section~\ref{sec:result}.
The LLP proper lifetimes for each mass point were also tuned to provide a set of samples with a mean lab-frame decay length of 9~m, and these samples are used to confirm the accuracy of the extrapolation method.
 
Multijet samples generated with \textsc{Pythia} 8.186
are used to determine the systematic uncertainties (described in Section~\ref{sec:syst}) in the displaced tracking and vertex reconstruction in the ID (described in Section~\ref{sec:reco}).
The A14 tune was used together with the NNPDF2.3LO PDF set.
 
The generated events for all MC samples described above were processed through a full simulation of the ATLAS detector geometry and response~\cite{SOFT-2010-01} using the \textsc{Geant4}~\cite{Agostinelli:2002hh} toolkit.
To model the effect of multiple $pp$ interactions per bunch crossing (pileup), additional simulated $pp$ interactions were overlaid onto each simulated hard-scatter event.
Pileup was simulated with \textsc{Pythia} 8.186 using the A2 set of tuned parameters and the MSTW2008LO~\cite{Ball:2014uwa} PDF set.
Per-event weights are applied to all simulated events such that the mean number of interactions per bunch crossing in simulation matches that in the data.
 
\section{Trigger}
\label{sec:trig}
 
Signal region events are selected with the Muon RoI Cluster trigger, which was developed to identify events with displaced hadronic decays outside the last active layer in the hadronic calorimeter~\cite{TRIG-2012-02}.
 
Hadronic decays after the end of the hadronic calorimeter and before the first trigger plane in the MS are characterized by multiple muon regions of interest\footnote{In the MS, L1 muon RoIs are $\Delta\eta \times \Delta\phi = 0.2 \times 0.2$ ($\Delta\eta \times \Delta\phi = 0.1 \times 0.1$) regions in the barrel (endcaps) identified as containing an interesting feature.}  (RoIs) around the LLP line of flight.
The trigger is seeded by an L1 trigger that searches for two RoIs in the MS, each of which is consistent with a particle with $\pt$ > 10~GeV.
At the HLT, the trigger requires clusters of muon RoIs, in which a cluster is defined as a $\Delta R = 0.4$ region containing at least three (four) muon RoIs in the barrel (endcap) of the MS.
 
To correct for the differences in efficiency of the trigger on events in data compared with events in MC samples, scale factors are used and are determined to be 1.13 $\pm$ 0.01 for the barrel and 1.04 $\pm$ 0.02 for the endcaps~\cite{EXOT-2017-05}.
The difference in the scale factors between the barrel and the endcaps derives from the differences in the trigger chambers used in the barrel (RPC) and the endcaps (TGC).
 
\section{Reconstruction}
\label{sec:reco}
 
To reconstruct the decay products and vertices of LLPs, dedicated reconstruction algorithms are used in both the ID and the MS.
In the ID, large-radius tracking (LRT)~\cite{ATL-PHYS-PUB-2017-014} is employed after standard tracking is completed in order to reconstruct those tracks that do not point to the IP, and a displaced vertex reconstruction algorithm~\cite{PERF-2011-08,SUSY-2016-08} draws on the combined collection of standard and large-radius tracks to form displaced ID vertices.
 
In the MS, LLPs that decay hadronically after the last layer of the hadronic calorimeter are likely to produce narrow, high-multiplicity jets; several times as many hits are expected to be associated with the LLP decays compared with those associated with a muon.
The standard algorithms in the MS are not optimized to operate in such dense environments.
A special vertex reconstruction algorithm is employed for the reconstruction of MS vertices, which is described in Ref.~\cite{PERF-2013-01}, and previously used in Refs.~\cite{EXOT-2013-12,EXOT-2017-05}.

\subsection{Reconstruction of standard jets}
\label{sec:reco_jet}
 
Jets are reconstructed in the electromagnetic and hadronic calorimeters from energy deposits in neighboring calorimeter cells.
Three-dimensional topological clusters of the cells containing energy significantly above a noise threshold~\cite{PERF-2016-04,PERF-2014-07} are used as input to the anti-$k_t$ jet algorithm~\cite{Cacciari:2008gp}.
Jets are reconstructed with an $R = 0.4$ radius parameter using the \textsc{FastJet} 2.4.3~\cite{Fastjet} software package.
Jets energies are calibrated using the procedure described in Ref.~\cite{PERF-2016-04}.
 
\subsection{Reconstruction of standard tracks}
\label{sec:reco_ST}
Tracks in the ID are reconstructed using the energy deposits, or hits, left by charged particles.
The standard ATLAS tracking algorithm reconstructs \textit{inside-out} tracks based on seeds made of three space points\footnote{
A space point is formed from a single measurement in the pixel detector or from a pair of measurements from the double layers in the SCT, with one measurement from a strip in the axial direction and the other from a strip in the stereo direction.}
in the pixel and SCT detectors~\cite{ATLAS-CONF-2010-072}.
A window search is performed based on the seeds, and track candidates are formed by inputting the hits in the window into a Kalman filter~\cite{Fruhwirth:1987fm}.
A track candidate must pass selection requirements on the track parameters and the constituent hits, as outlined in Table~\ref{table:STvsLRT}.
Track candidates are passed through an ambiguity solver~\cite{ATLAS-CONF-2010-072} which evaluates the track candidates by using the hits, as well as the $\chi^2$ of the track fit.
Successful track candidates are extended into the TRT, and tracks are kept whether or not the TRT extension is successful.
 
After the completion of the inside-out tracking pass, an \textit{outside-in} tracking pass is performed.
Standalone TRT segments are created, seeded by energy deposits in the electromagnetic calorimeter.
The standalone TRT segments are extended back into the SCT and pixel detectors, using hits that were not included in tracks reconstructed in the inside-out tracking.
Standalone TRT segments that fail the extension into the silicon detectors are retained.
Outside-in tracks must also pass the track impact parameter requirements in Table~\ref{table:STvsLRT}.
 
\subsection{Reconstruction of large-radius tracks}
\label{sec:reco_LRT}
 
The standard ATLAS tracking procedure is optimized for the reconstruction of tracks that originate very close to the IP and has strict restrictions on the impact parameters of reconstructed tracks to reduce the reconstruction of fake tracks.
Tracks produced in displaced decays often have impact parameters that are larger than the maximum impact parameter allowed by the standard tracking reconstruction algorithm.
In order to reconstruct these tracks, LRT is performed, taking as inputs the hits that are left over from the standard tracking.
This iteration is performed with loosened requirements on the transverse ($d_0$) and longitudinal ($z_0$) track impact parameters, outlined in Table~\ref{table:STvsLRT}, in order to provide increased efficiency for displaced tracks.
 
\begin{table}[htb!]
\centering
\begin{threeparttable}
\caption{Track parameter requirements for inside-out standard and large-radius tracks.}
\begin{tabular}{ l c c }
\hline
Track parameter 				& Standard	& Large radius \\
\hline
Maximum $|d_0|$ 				& 10~mm		& 300~mm  \\
Maximum $|z_0|$				& 250~mm 	& 1500~mm  \\
Minimum  \pt  					& 400~\MeV 	& 500~\MeV  \\
Maximum track $|\eta|$			& 2.7 		& 5.0 \\
Minimum silicon hits				& 7 			& 7 \\
Minimum unshared silicon hits 		& 6 			& 5 \\
\hline
\end{tabular}
\label{table:STvsLRT}
\end{threeparttable}
\end{table}
 
Additionally, in order to increase the efficiency of the reconstruction of the displaced decays, the requirement on the maximum number of unshared hits is relaxed.
 
\subsection{Displaced vertex reconstruction in the MS}
\label{sec:reco_MSVX}
 
The dedicated MSVx reconstruction algorithm makes use of the two multilayers (ML) in each MDT chamber~\cite{PERF-2013-01}.
Straight-line segments are created from at least three MDT hits in each of the MLs using a minimum $\chi^2$ fit.
The segments from the two MLs are matched to form \textit{tracklets}, which are used to reconstruct the MSVx positions.
The vertex reconstruction algorithm proceeds separately in the barrel and the endcaps of the MS, extrapolating the tracklets through the magnetic field that is present in the barrel, and extrapolating tracklets as straight lines in the endcaps where the MDT chambers are not immersed in magnetic field.
In the barrel, the $\chi^2$ probability of the vertex is required to be $\geq 5$\%.
If it is less than 5\%, the tracklet with the largest contribution to the $\chi^2$ is removed and the vertex is re-fit.
This process is repeated until the $\chi^2$ probability of the vertex is $\geq 5$\% or the vertex has fewer than three tracklets.
In the endcap, all tracklets in the vertex must be within 30~cm of the calculated vertex position, otherwise the farthest tracklet from the vertex is removed and the vertex is re-fit.
In both the barrel and the endcaps, each MSVx is required to have at least three tracklets.
 
\subsection{Displaced vertex reconstruction in the ID}
\label{sec:reco_IDVX}
 
A displaced vertex reconstruction algorithm is used to reconstruct the LLP decays in the ID.
The algorithm takes both the standard and large-radius tracks as input and selects tracks that meet the criteria listed in the upper section of Table~\ref{table:trackreqs}.
 
\begin{table}[htb!]
\centering
\begin{threeparttable}
\caption{Track parameter requirements for the reconstruction of displaced vertices in the ID.}
\begin{tabular}{ p{5cm}  c }
\hline
Track parameter & Requirement \\
\hline
Track $|d_0|$	& 2~mm < $|d_0|$ < 300~mm	\\
Track $|z_0|$ 	& < 1500~mm	\\
Track $\pt$ 	& > 1~\GeV	\\
Number of SCT hits 			& $\geq 2$	\\
Number of pixel and TRT hits 	& $n_{\mathrm{pixel}} \geq 2$ or $n_{\mathrm{TRT}} > 0$ \\
\hline
\hline
\multicolumn{2}{c}{No hits on track may be present before the vertex} \\
\multicolumn{2}{c}{Hits on track must be present in the layer following the vertex} \\
\hline
\end{tabular}
\label{table:trackreqs}
\end{threeparttable}
\end{table}
 
From this collection, a set of two-track vertices is formed from all possible pairs of intersecting tracks.
The tracks making up these two-track vertices are required not to have any hits in pixel or SCT layers at smaller radii than the vertex position, and must have hits on the next possible pixel or SCT layer, unless the vertex is within a few mm of the layer.
The two-track vertex seeds are merged into multi-track vertices if they are within $d/\sigma_d < 3$ (where $d$ is the distance between the two-track vertices and $\sigma_d$ is the uncertainty in the distance).
Poorly associated tracks, with track $\chi^2 > 6$, are removed from the multi-track vertices and the vertices are re-fit.
This process is repeated until there are no more pairs of vertices satisfying $d/\sigma_d < 3$.
In the final step, all vertices within 1~mm are merged and the vertex fit is recalculated.
 
\section{Event selection}
\label{sec:select}
 
All events used in the analysis are required to contain a primary vertex (PV), associated with the $pp$ hard scatter~\cite{ATL-PHYS-PUB-2015-026}.
The PV must have at least two tracks, each with $\pt > 400 \, \MeV$.
If more than one vertex exists satisfying these criteria, the PV is chosen as the vertex with the largest sum of the squares of the \pt~of all tracks associated with the vertex.  
The events must pass the Muon RoI Cluster trigger, and pass the veto on the triggers from the CR analysis~\cite{CalRatio2016}.
The events must contain a good MSVx (described in Section~\ref{sec:MSVxSel}) matched within  $\Delta R < 0.4$ to the triggering muon cluster.
Finally, events are required to have a good IDVx (described in Section~\ref{sec:IDVxSel}), and the MSVx and IDVx must have an angular separation of $\Delta R> 0.4$.\footnote{The separation is measured from the axis of the MSVx, which is defined with respect to the detector coordinate system, and from the ($\eta$, $\phi$) of the tracks associated with the IDVx.}

\subsection{MS vertex selection}
\label{sec:MSVxSel}
The primary source of background that mimics LLP decays in the MS is jets that punch through the calorimeter.
In order to reduce the background from these punch-through jets, each MSVx is required to pass certain isolation requirements developed in the MS analysis~\cite{EXOT-2017-05}.
The MS vertices are required to be isolated by $\Delta R > 0.3$ ($\Delta R > 0.6$) in the barrel (endcaps) from jets with $p_{\mathrm{T}} > 30 \, \GeV$ that are matched to the PV using a jet vertex tagger discriminant~\cite{PERF-2014-03} and have $\log_{10}(E_{\mathrm{HAD}}/E_{\mathrm{EM}}) < 0.5$.\footnote{The term $\log_{10}(E_{\mathrm{HAD}}/E_{\mathrm{EM}})$, quantifies the ratio of the energy deposited in the hadronic calorimeter to the energy deposited in the electromagnetic calorimeter.
The purpose of this requirement is to prevent highly displaced jets from being used in the isolation, because LLPs that decay at the end of the hadronic calorimeter may leave a jet and also create a decay vertex in the MS.}
Isolation from these jets also reduces the contamination from multijet events.
To further reduce the background from multijet events, each MSVx is required to be isolated from activity in the ID.
The vector sum of the transverse momenta of tracks in a  $\Delta R = 0.2$ cone around the MSVx is required to be $\Sigma \pt$ < 10 \GeV, and in the barrel (endcaps) the MSVx must be $\Delta R > 0.3$ ($\Delta R > 0.6$) from any tracks with $\pt > 5 \, \GeV$.
The tracks used for this isolation must point back to the PV.
Additionally, they are required to have at least seven silicon hits and no shared silicon hits, or at least ten silicon hits.
To reduce the contribution from electronic noise, cosmic-ray muons, and machine-induced background, each MSVx is also required to have a minimum number of hits associated with the vertex in the MDT ($n_{\mathrm{MDT}}$), RPC ($n_{\mathrm{RPC}}$), and TGC ($n_{\mathrm{TGC}}$), outlined in Table~\ref{table:goodMSVx}.
A maximum $n_{\mathrm{MDT}}$ selection is applied to reduce the background from coherent noise bursts in the MDTs.
 
MS vertices are reconstructed by different algorithms in the barrel and the endcaps.
When an LLP decays in the barrel-endcap transition region, $0.8 < |\eta| < 1.3$, the resulting hits will be split between the two algorithms.
This leads to a low reconstruction efficiency in the transition region, since neither algorithm has access to all hits, although occasionally two separate vertices will be reconstructed from a single LLP decay.
The barrel-endcap transition region overlaps very closely with the barrel-endcap transition region in the hadronic calorimeter, $0.7 < |\eta| < 1.2$, in which the probability of punch-through jets is higher.
Thus, each MSVx is required to be contained in the barrel with $|\eta_{\mathrm{MSVx}}| < 0.7$ or in the endcap with $|\eta_{\mathrm{MSVx}}| > 1.3$ to remove vertices in the barrel–endcap transition region of the MS and hadronic calorimeter.
A vertex that meets all of the necessary criteria is considered to be a good MSVx.
 
\begin{table}[!htb]
\centering
\begin{threeparttable}
\caption{MSVx selection requirements.}
\begin{tabular}{ l  c  c   }
\hline
& Barrel 						& Endcaps						\\
\hline
Precision chamber hits								& \multicolumn{2}{c}{$300 < n_{\mathrm{MDT}} < 3000$} 					\\
Trigger chamber hits									& $n_{\mathrm{RPC}} > 250$		& $n_{\mathrm{TGC}} > 250$ 			\\
Isolation from \pt > 5 \GeV~tracks							& $\Delta R > 0.3$				& $\Delta R > 0.6$ 				\\
$\Sigma \pt$ of tracks in $\Delta R = 0.2$ cone 					& < 10~\GeV					& < 10~\GeV					\\
Isolation from \pt > 30 \GeV~jets 							& $\Delta R > 0.3$				& $\Delta R > 0.6$ 				\\
MSVx $|\eta|$										& $|\eta_{\mathrm{MSVx}}|$ < 0.7	& 1.3 < $|\eta_{\mathrm{MSVx}}|$ < 2.5	\\
\hline
 
\end{tabular}
\label{table:goodMSVx}
\end{threeparttable}
\end{table}
 
\subsection{ID vertex selection}
\label{sec:IDVxSel}
One of the primary sources of background for a search for displaced hadronic decays in the ID is vertices from interactions between particles and layers of detector material.
Such hadronic interactions may result in reconstructed vertices that are indistinguishable from the reconstructed vertices of signal decays in the same region of space.
In order to remove this source of background, a map is created using displaced vertices found in minimum-bias data in which the decays from known long-lived hadrons have been removed, as described in Refs.~\cite{PERF-2015-06,PERF-2015-07}.
This map is used to create a material veto, which removes vertices in regions of space that were found to contain material.
The fiducial volume defined by IDVx, $R = \sqrt{x^{2} + y^{2}}$ < 300~mm and $|z|$ < 300~mm is reduced by 42\% due to this material veto~\cite{SUSY-2016-08}; however, the material veto reduces the number of background events by more than a factor of 50.
Some modules in the ATLAS detector are occasionally disabled. 
To mimic this effect in simulation, a disabled-module veto is employed that accounts for the fact that a disabled module could cause a track not to have a hit immediately after a vertex, causing the vertex to be rejected.
This disabled-module veto operates by removing vertices in regions immediately before the location of the disabled modules.
The disabled-module veto leads to a minor loss in fiducial volume of 2.3\%~\cite{SUSY-2016-08}, and is applied in both data and simulation.
 
To reject poorly reconstructed vertices resulting from random track crossings, the $\chi^2$ value of the vertex fit divided by the number of degrees of freedom is required to be less than 5.
 
In addition to the IDVx reconstruction requirement that the tracks forming the vertices have $|d_0| > 2$~mm, a minimum radial distance of 4~mm is imposed between the IDVx and the reconstructed PV to further reduce the background contribution from $b$-hadrons.
 
ID vertices are required to have a separation of $\Delta R > 0.4$ from the nearest selected MSVx in the event in order to reduce the probability that one high-energy hadronic jet could cause a background vertex simultaneously in the ID and in the MS.
 
\begin{figure}[!htb]
\centering
\subfloat[]{\label{fig:vtxnTrks}\includegraphics[width=0.495\textwidth]{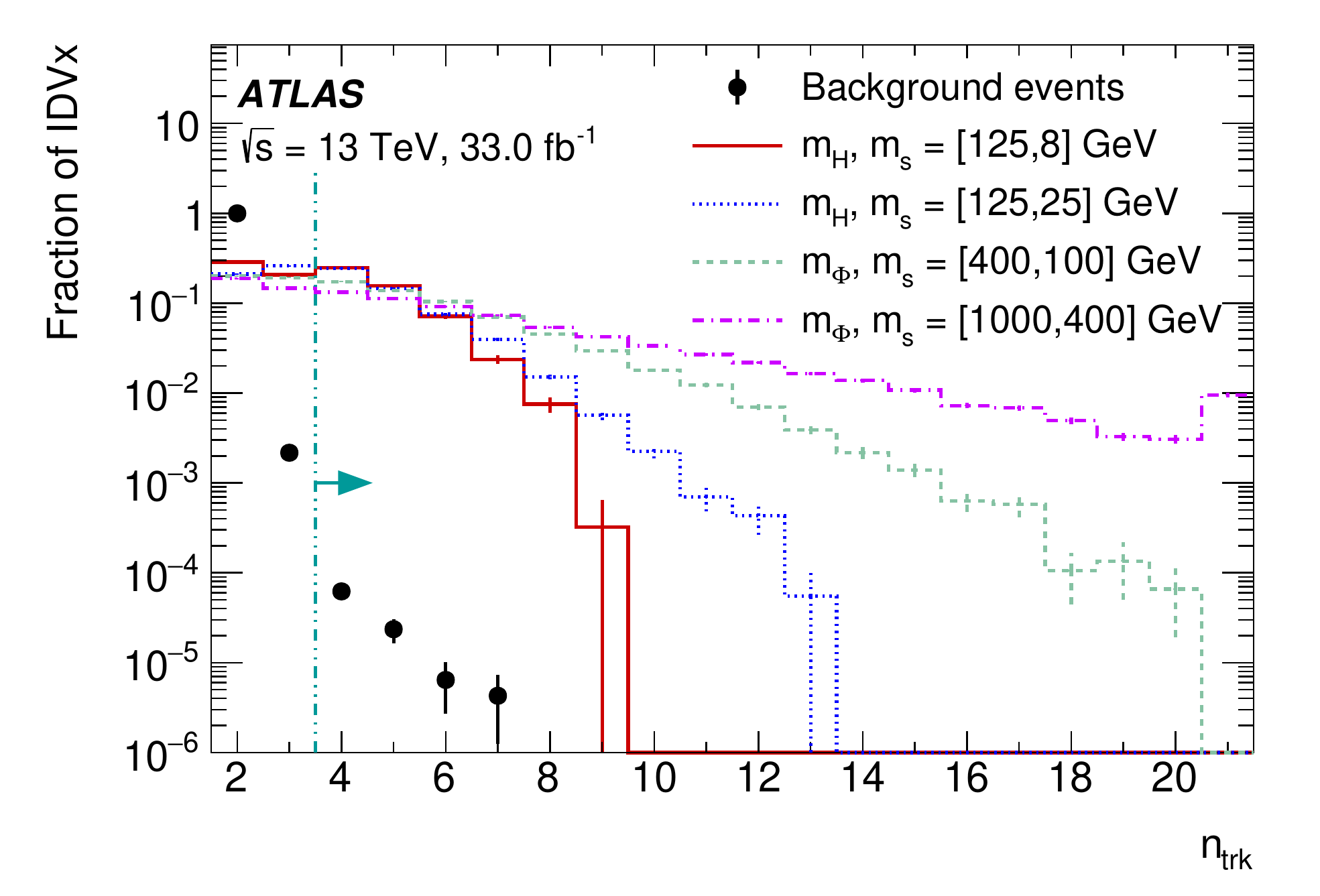}}
\,
\subfloat[]{\label{fig:vtxMass}\includegraphics[width=0.495\textwidth]{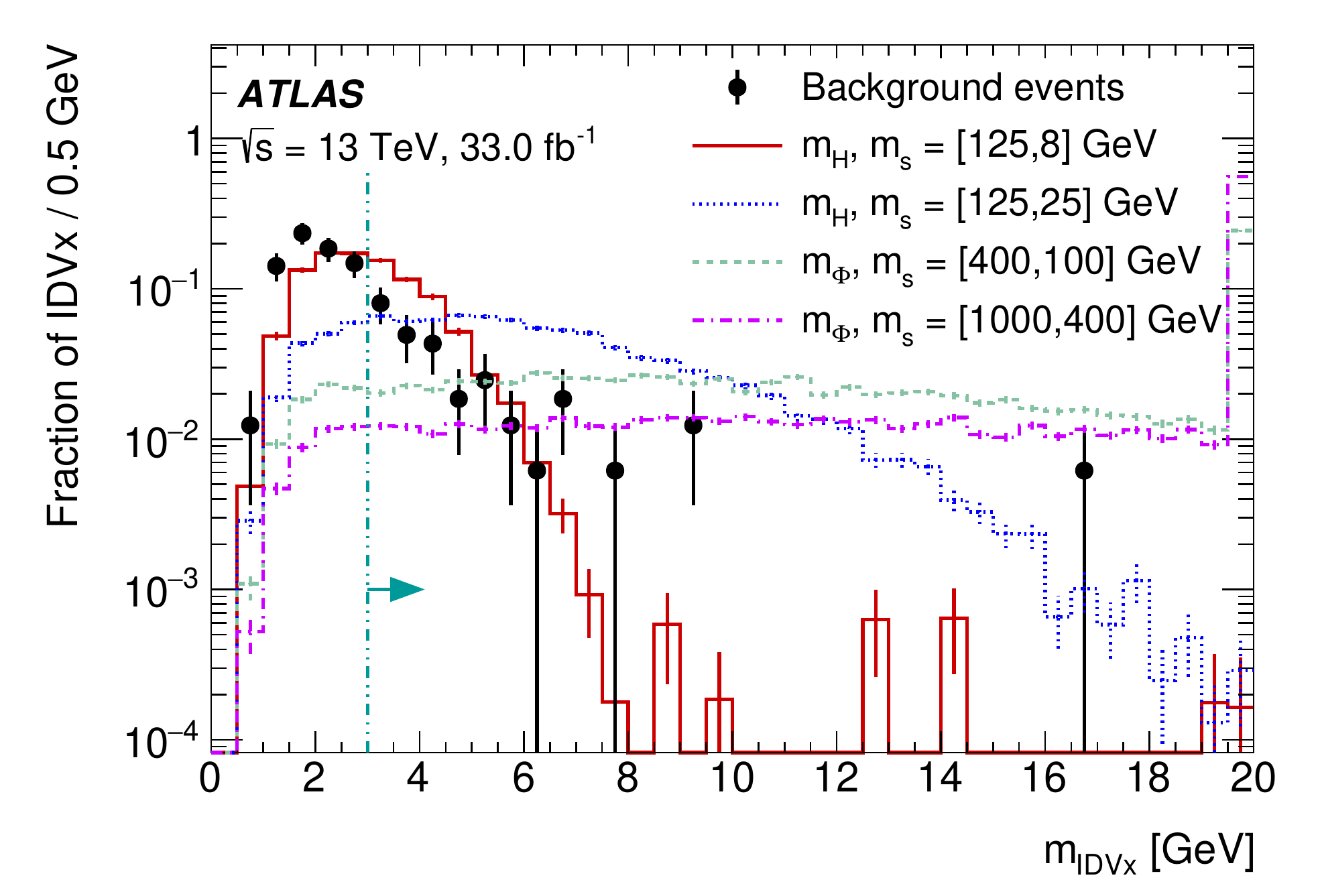}}
\caption{Distributions of (a) $n_{\mathrm{trk}}$ and (b) $m_{\mathrm{IDVx}}$ for simulated signal MC and data background samples.
The vertices shown are required to pass all selection requirements outlined in Table~\ref{table:IDVxreqs} except the one displayed in the plot.
The dashed lines and arrows indicate the selection requirement on the parameter shown.
The last bin includes overflow.
}
\label{fig:SvB_massnTrks}
\end{figure}

The number of charged decay products from a hadronically decaying LLP is expected to be much higher than from vertices constructed from fake tracks or from random crossings of tracks.
Thus, the number of tracks associated with the IDVx is an important discriminant between background and signal vertices.
Figure~\ref{fig:vtxnTrks} shows the distributions of the number of tracks associated with each IDVx ($n_{\mathrm{trk}}$) for reconstructed vertices in signal MC samples compared with those reconstructed in data.
In Figure~\ref{fig:SvB_massnTrks}, the signal MC events used have no additional selection requirements in order to reduce the statistical uncertainty.
The background events in data are those events in the $Bkg$ region of the background estimation method described in Section~\ref{sec:background}.
The reconstructed vertices in the signal MC samples are required to be $dr = \sqrt{dx^2 + dy^2 + dz^2}< $ 5~mm of the LLP decay position, and have at least two tracks matched\footnote{A reconstructed track is considered matched to a particle if it has $p > 0.5$, where $p$ represents the weighted fraction of hits in the reconstructed track that are associated with the generated particle. The hits are assigned a weight based on the sub-detector in which they were found, in order to reflect the intrinsic resolution of the measurement in that sub-detector.} to particles produced in the LLP decay.
The reconstructed IDVx distributions in the background events in data are dominated by vertices with $n_{\mathrm{trk}} = 2$, and when two-track vertices are removed, by $n_{\mathrm{trk}} = 3$.
For an IDVx to be considered in the signal region, it is required to have $n_{\mathrm{trk}} \geq 4$.

Each IDVx is also required to pass a minimum vertex mass ($m_{\mathrm{IDVx}}$) selection of 3~\GeV, where the invariant vertex mass is computed assuming that the tracks originate from charged pions.
Figure~\ref{fig:vtxMass} shows the distributions of $m_{\mathrm{IDVx}}$ in background events and several signal samples.
The selection requirement of $m_{\mathrm{IDVx}} > 3~$\GeV~has the greatest impact on the selection efficiency for the lowest-mass LLPs, removing approximately half of the reconstructed ID vertices in the $m_{H}, m_{s} = [125,8]$~\GeV~signal MC sample that pass the other selection requirements listed in Table~\ref{table:IDVxreqs}.
The heavier the mass of the LLP, the less impact the selection $m_{\mathrm{IDVx}} > 3~$\GeV~has on the IDVx selection efficiency.
For the $m_{\Phi}, m_{s} = [1000,400]$~\GeV~signal MC sample, this selection removes less than 4\% of the reconstructed ID vertices that pass the other signal selections.
In data, the selection on $m_{\mathrm{IDVx}}$ removes 70\% of reconstructed ID vertices which pass all the other selections.
 
The requirements on the signal region IDVx are summarized in Table~\ref{table:IDVxreqs}.
 
\begin{table}[htb!]
\centering
\begin{threeparttable}
\caption{IDVx selection requirements.}
\begin{tabular}{ p{5cm} c }
\hline
Vertex parameter				& Requirement				\\
\hline
Vertex $R$					& < 300~mm				\\
Vertex $|z|$					& < 300~mm				\\
IDVx position					& Pass material veto			 \\
IDVx position					& Pass disabled-module veto	\\
Radius from PV					& > 4~mm					\\
$\chi^2 / n_{\mathrm{DoF}}$		& < 5				\\
$\Delta R$ from nearest good MSVx	&  $>$ 0.4 				\\
$m_{\mathrm{IDVx}}$			& > 3~\GeV			\\
$n_{\mathrm{trk}}$				& $\geq 4$ 				\\
\hline
\end{tabular}
\label{table:IDVxreqs}
\end{threeparttable}
\end{table}
 
\begin{figure}[!htb]
\centering
\subfloat[]{\label{subfig:VtxRecoEff_Higgs}\includegraphics[width=0.495\textwidth]{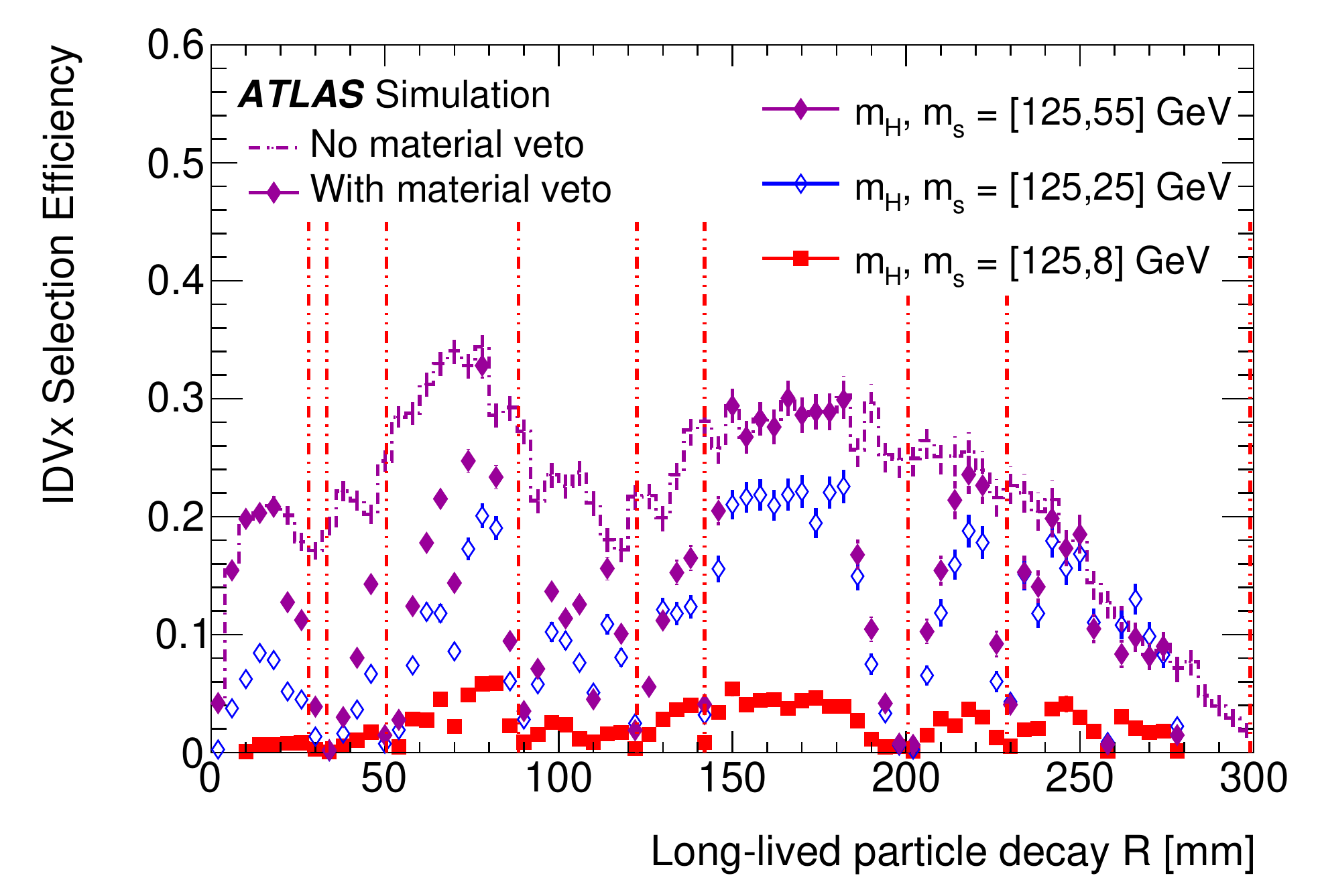}}
\,
\subfloat[]{\label{subfig:VtxRecoEff_LLP50}\includegraphics[width=0.495\textwidth]{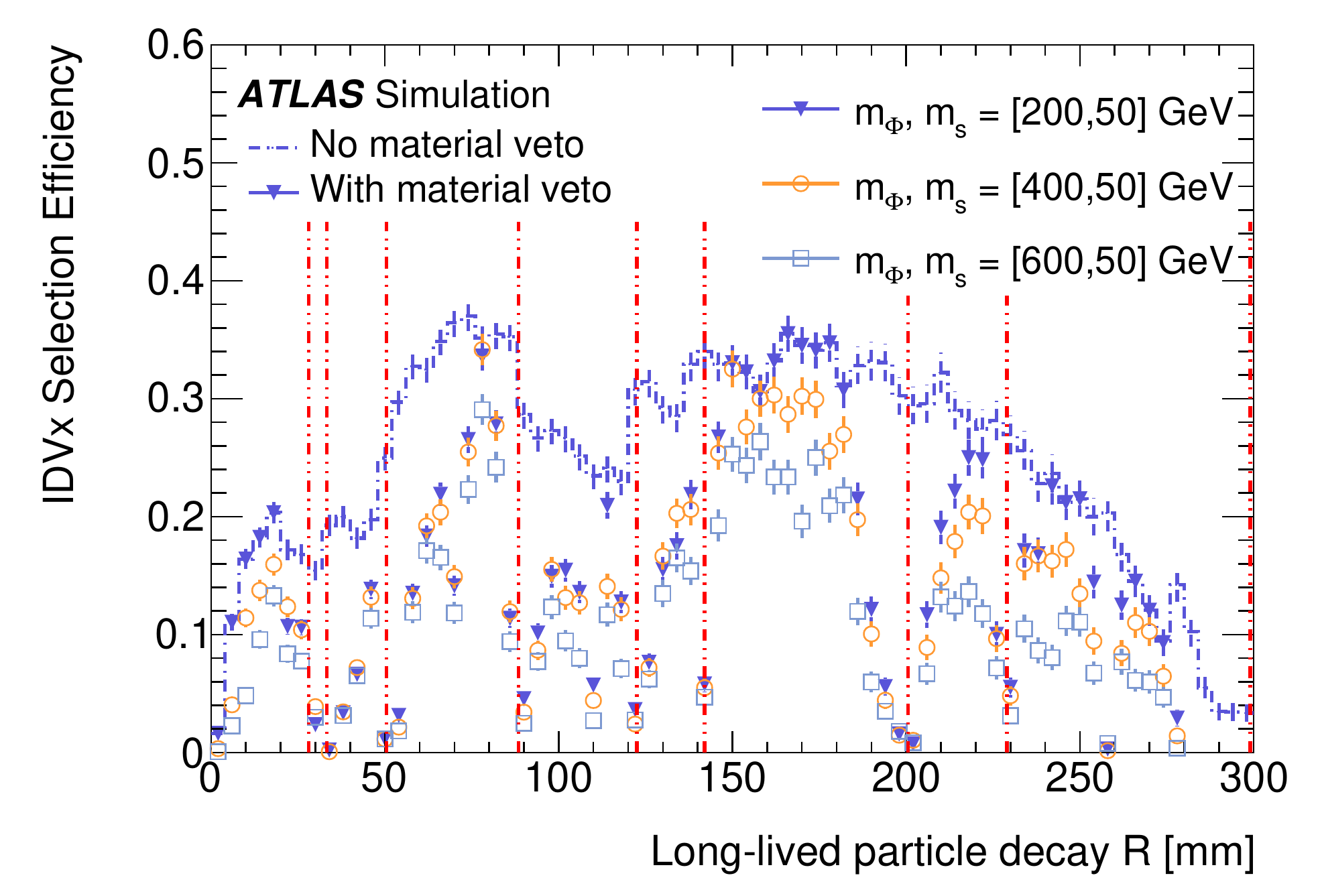}}
\caption{IDVx selection efficiency as a function of the radial decay position for several (a) 125~\GeV~$H \rightarrow ss$ and (b) $\Phi \rightarrow ss$ mass points with $m_s = 50$~\GeV.
The dashed vertical red lines indicate the main radial material layers in the inner detector that are included in the material veto.
In each figure, the IDVx selection efficiency for one signal MC sample is shown before and after the requirement of the material veto to demonstrate the impact it has on the IDVx selection efficiency.}
 
\label{fig:VtxRecoEff}
\end{figure}

Figure~\ref{fig:VtxRecoEff} shows the IDVx selection efficiency, including the reconstruction efficiency, as a function of the decay radius of the LLP for several mass points.
The efficiency is defined as the fraction of LLP decay vertices in the fiducial volume $R$, $|z| < 300$~mm that are within 5~mm of reconstructed vertices that have at least two tracks matched to the decay products from the LLP and meet all  the selection criteria listed in Table~\ref{table:IDVxreqs}.
 
Figure~\ref{subfig:VtxRecoEff_Higgs} shows that for a fixed mediator mass the IDVx selection efficiency increases with increasing LLP mass due to the larger impact of the selection on $m_{\mathrm{IDVx}}$ and $n_{\mathrm{trk}}$ at smaller LLP masses.
For a fixed LLP mass, a higher boost leads to a decreased IDVx selection efficiency (Figure~\ref{subfig:VtxRecoEff_LLP50}) because decay products that point back to the IP are not included in the displaced vertex reconstruction, due to the selection of $|d_0| > 2$~mm on associated tracks.
 
The structure in the IDVx selection efficiency as a function of the LLP decay radius $R$ is primarily due to the impact of the material veto, as shown in Figures~\ref{subfig:VtxRecoEff_Higgs} and~\ref{subfig:VtxRecoEff_LLP50} for one mass point by the inclusion of the selection efficiency without the material veto.
The shape versus the LLP decay radius is also impacted by other factors such as the hit requirements in the track creation and vertex reconstruction and the selection requirements on the constituent track $|d_0|$ and the distance from the PV.

\subsection{Overall selection efficiency}
 
Table~\ref{table:cutFlowPercentage} details the total efficiency after each selection requirement applied to signal events, as well as the relative efficiency of each subsequent selection requirement.
All events in signal MC samples are found to contain a PV, thus this selection requirement is not included in the table.
Many of the signal MC events generated have LLPs that decay outside of the fiducial volumes of the ID and the MS, and there is no fiducial requirement on events displayed in this table.
Requiring events to contain one LLP decaying in the fiducial volume of the MS and one in the fiducial volume of the ID increases the relative efficiency of every selection requirement.
It is apparent from Table~\ref{table:cutFlowPercentage} that the relative efficiency of all IDVx-related selection requirements increases with increased LLP mass, but decreases with increased LLP boost (this can also be seen in Figure~\ref{fig:VtxRecoEff}).
Events with higher LLP mass are also more likely to pass the trigger requirements and contain a good MSVx.
The boost of the LLP is associated with a higher probability to pass the trigger requirements but a lower probability to contain a good MSVx, for events that pass the trigger.
 
\begin{table}[!htb]
\caption{Total and relative efficiency for each selection requirement for the several signal MC mass points.
Here, `Pass trigger' refers to passing the Muon RoI Cluster trigger and passing the veto on the CR triggers, `Good MSVx' includes all the MSVx selection requirements described in Table~\ref{table:goodMSVx} as well as being matched to the muon RoI cluster and `IDVx' includes all selection requirements in Table~\ref{table:IDVxreqs} except those on $n_{\mathrm{trk}}$ and $m_{\mathrm{IDVx}}$, which are listed separately.
All efficiencies are computed using signal MC samples with a mean lab-frame decay length of 5~m, the proper lifetime of each mass point is listed in the table.}
\centering
\begin{tabular}{  c c  |  l  c c  c  c  c }
\hline
Selection requirements					&  					&  \multirow{2}{*}{Efficiency}	&Pass	& Good 	& \multirow{2}{*}{IDVx}	&  $n_{\mathrm{trk}}$ 	& $m_{\mathrm{IDVx}}$	\\
Mass point [\GeV ] 						& $c\tau$ [m]			&  						&trigger	& MSVx	& 					&   $\geq$ 4 			&  $>$ 3 \GeV	\\
\hline
\multirow{2}{*}{$m_{H},m_s = [125,8]$}		&  \multirow{2}{*}{0.200}	& Total					& 2.71\%	& 1.07\%	& 0.13\%				& 0.005\%				& 0.003\%	\\
& 					& Relative					& 2.71\%	& 39.3\%	& 12.5\%				& 3.61\%				& 63.2\%		\\
\hline
\multirow{2}{*}{$m_{H},m_s = [125,25]$}		& \multirow{2}{*}{0.760}	& Total					& 5.13\%	& 2.23\%	& 0.30\%				& 0.03\%				& 0.02\%		\\
&					& Relative					& 5.13\%	& 43.5\%	& 13.3\%				& 9.15\%				& 81.1\%		\\
\hline
\multirow{2}{*}{$m_{H},m_s = [125,55]$}		& \multirow{2}{*}{1.540}	& Total					& 1.98\%	& 0.75\%	& 0.11\%				& 0.01\%				& 0.01\%		\\
&					& Relative					& 1.98\%	& 37.9\%	& 14.2\%				& 10.1\%				& 85.4\%		\\
\hline
\multirow{2}{*}{$m_{\Phi},m_s = [200,50]$}	& \multirow{2}{*}{1.070}	& Total					& 7.06\%	& 3.05\%	& 0.47\%				& 0.07\%				& 0.06\%		\\
&					& Relative					& 7.06\%	& 43.2\%	& 15.3\%				& 15.0\%				& 83.9\%		\\
\hline
\multirow{2}{*}{$m_{\Phi},m_s = [400,50]$}	& \multirow{2}{*}{0.700}	& Total					& 13.7\%	& 5.02\%	& 0.73\%				& 0.10\%				& 0.09\%		\\
&					& Relative					& 13.7\%	& 36.5\%	& 14.5\%				& 14.3\%				& 83.5\%		\\
\hline
\multirow{2}{*}{$m_{\Phi},m_s = [600,50]$}	& \multirow{2}{*}{0.520}	& Total					& 16.4\%	& 4.77\%	& 0.69\%				& 0.08\%				& 0.07\%		\\
&					& Relative					& 16.4\%	& 29.0\%	& 14.5\%				& 12.2\%				& 78.4\%		\\
\hline
 
\end{tabular}
\label{table:cutFlowPercentage}
\end{table}
 
 
\section{Background estimation}
\label{sec:background}
 
Sources of background for ID vertices include reconstructed vertices from the interactions of particles with detector material, vertices created from fake tracks, and vertices from random track crossings.
The incidence of vertices created from fake tracks and from random crossings is correlated with the jet activity in the events; events with a greater number of jets are more likely to include vertices from background.
The background vertices are predominantly removed by the IDVx selection.
A data-driven background estimation method is employed to determine the residual contribution to the events in the signal region from ID vertices from all sources of background.
 
The number of background events in the signal region is estimated by defining a set of background events that are designed to be approximately free of signal contamination and determining the fraction of those events that contain an IDVx that passes the full IDVx selection.
This fraction is applied to events that pass the full event selection except for the requirement of an IDVx, to estimate how many background events contain an MSVx and an IDVx.
 
\begin{table}[!htb]
\centering
\caption{
The data events used in the background estimation.
These events include the background events, whose selection is defined in the text, divided into all background events ($\mathit{Bkg}$), and those that contain at least one IDVx that passes the full IDVx requirements ($\mathit{Bkg}$+$\mathit{IDVx}$).
The other events making up the plane are the signal region events ($\mathit{Sig}$), and events that pass all signal region requirements except for the inclusion of an IDVx ($\mathit{Sig}$--$\mathit{IDVx}$).
}
\begin{tabular}{ c | c | c  }
\hline
& Background		& Muon RoI Cluster trigger	\\
& events			& events with a good MSVx	\\
\hline
Has IDVx passing full signal selection	& $\mathit{Bkg}$+$\mathit{IDVx}$		&  $\mathit{Sig}$			\\
\hline
Agnostic to IDVx  					& $\mathit{Bkg}$			& $\mathit{Sig}$--$\mathit{IDVx}$				\\
\hline
\end{tabular}
\label{fig:ABCD}
\end{table}
 
The selection of background events is designed to limit the possibility of signal contamination.
The background events ($\mathit{Bkg}$), shown in the lower left in Table~\ref{fig:ABCD}, are required to pass the single-muon trigger described in Section~\ref{sec:data} and are required to pass a veto on the Muon RoI Cluster trigger used to collect events in the signal region.
The background events are also required to contain two isolated muons with $\pt > 25 \, \GeV$ and $\pt > 20 \, \GeV$.
When applied to events in the signal MC samples, the requirements used to define the $\mathit{Bkg}$ events select fewer than 0.1\% of events for any given mass point, without including the requirement of reconstructing and selecting both an MSVx and an IDVx.
Accounting for pre-existing limits on the branching ratio and cross section for any given mass point, signal is expected to make up less than 0.005\% of the events populating the $\mathit{Bkg}$ region.
 
The selection for the $\mathit{Bkg}$ events is agnostic to the presence of an IDVx.
The $\mathit{Bkg}$+$\mathit{IDVx}$ events, shown in the upper left in Table~\ref{fig:ABCD}, are required to pass the same trigger and isolated muon background requirements as the $\mathit{Bkg}$ events, but are also required to contain at least one IDVx that passes all of the IDVx selections outlined in Table~\ref{table:IDVxreqs}.
The number of events in regions $\mathit{Bkg}$+$\mathit{IDVx}$ and $\mathit{Bkg}$ ($N_{\mathit{Bkg+IDVx}}$ and $N_{\mathit{Bkg}}$) are used to calculated a factor $F = N_{\mathit{Bkg+IDVx}}/N_{\mathit{Bkg}}$, which represents the probability that a given event will contain an IDVx from background that meets the selection criteria.
 
Signal region events ($\mathit{Sig}$) in the upper right of Table~\ref{fig:ABCD} are those which pass the full signal selection described in Section~\ref{sec:select}.
Events that pass the full signal selection but are not required to contain an IDVx ($\mathit{Sig}$--$\mathit{IDVx}$) are in the lower right of Table~\ref{fig:ABCD}.
The number of events in the signal region ($N_{\mathit{Sig}}$) which contain an IDVx from background can be estimated using the number of $\mathit{Sig}$--$\mathit{IDVx}$ events ($N_{\mathit{Sig-IDVx}}$) and the factor $F$ defined above.
Hence, the number of $\mathit{Sig}$ region events which are expected to contain an IDVx from background is estimated as $N_{\mathit{Sig}}^{\mathrm{bkg \, pred.}} = N_{\mathit{Sig-IDVx}}\times F = N_{\mathit{Sig-IDVx}} \times N_{\mathit{Bkg+IDVx}} / N_{\mathit{Bkg}}$.

There are 45 events in the $\mathit{Bkg}$+$\mathit{IDVx}$ region out of 6,099,660 events in the $\mathit{Bkg}$ region, giving a factor $F = N_{\mathit{Bkg+IDVx}}/N_{\mathit{Bkg}} = (7.4 \pm 1.1 \, (\mathrm{stat.}))\times10^{-6}$.
The $\mathit{Sig}$--$\mathit{IDVx}$ region contains 156,805 events; the predicted number of background events in the signal region is then $N_{\mathit{Sig}}^{\mathrm{pred}} =N_{\mathit{Sig-IDVx}} \times F = N_{\mathit{Sig-IDVx}} \times N_{\mathit{Bkg+IDVx}}/ N_{\mathit{Bkg}} = 1.16 \pm 0.18 \, (\mathrm{stat.}) $.
 
The validation of the background estimation is performed using two sets of validation regions, in which the predicted and observed numbers of events are compared in regions containing vertices similar to those passing the IDVx signal selection (Table~\ref{fig:ABCD_CRboth}).
This validation serves as a cross-check that the fraction of background events that contain an IDVx is not significantly different from the fraction of events passing the Muon RoI Cluster trigger, with or without a selected MSVx, that contain an IDVx from background.
 
The first set of validation regions, the $\mathit{Bkg},2\mathit{\mbox{-}trk}$ and $\mathit{Val},2\mathit{\mbox{-}trk}$ regions, are similar to the $\mathit{Bkg}$+$\mathit{IDVx}$ and $\mathit{Sig}$ regions except that instead of containing an IDVx that meets the full IDVx selection criteria, the selection on $n_{\mathrm{trk}}$ is changed from $n_{\mathrm{trk}} \geq 4$ to $n_{\mathrm{trk}} = 2$.
Vertices with $n_{\mathrm{trk}} = 2$ are chosen because the IDVx distribution in background events is dominated by two-track vertices, as shown in Figure~\ref{fig:vtxnTrks}, and the signal contamination is small.
This allows the fraction of vertices to be examined in events that otherwise pass all signal region selections.
 
\begin{table}[!htb]
\centering
\caption{The events used for the validation of the background estimation, alongside the events used for the background estimate.
The $\mathit{Bkg},2\mathit{\mbox{-}trk}$ and $\mathit{Val},2\mathit{\mbox{-}trk}$ validation regions contain ID vertices that have $n_{\mathrm{trk}} = 2$.
The $\mathit{Trig},3\mathit{\mbox{-}trk}$ and $\mathit{Trig}$ validation regions are events that pass the Muon RoI Cluster trigger but are agnostic to the presence of MS vertices.
The $\mathit{Bkg},3\mathit{\mbox{-}trk}$ and $\mathit{Trig},3\mathit{\mbox{-}trk}$ validation regions contain ID vertices with $1~\GeV < m_{\mathrm{IDVx}} < 3~\GeV$~and $n_{\mathrm{trk}} = 3$.
}
\begin{tabular}{  c |  c | c | c  }
\hline
& Background			& Muon RoI Cluster,		& Muon RoI Cluster,				\\ 
& events				& agnostic to MSVxs		& with a good MSVx				\\ 
\hline
Has IDVx, $n_{\mathrm{trk}} \geq$ 4, $m_{\mathrm{IDVx}}  > 3~\GeV$	& $\mathit{Bkg}$+$\mathit{IDVx}$			&					&  $\mathit{Sig}$				\\ 
\hline
Has IDVx, $n_{\mathrm{trk}} = 3$, $1 <  m_{\mathrm{IDVx}} < 3~\GeV$	&  $\mathit{Bkg},3\mathit{\mbox{-}trk}$	& $\mathit{Trig},3\mathit{\mbox{-}trk}$	&							\\ 
\hline
Has IDVx, $n_{\mathrm{trk}}$ = 2, $m_{\mathrm{IDVx}}$ > 3~\GeV		& $\mathit{Bkg},2\mathit{\mbox{-}trk}$ 	&					& $\mathit{Val},2\mathit{\mbox{-}trk}$				\\ 
\hline
Agnostic to IDVx											& $\mathit{Bkg}$ 				& $\mathit{Trig}$				&  $\mathit{Sig}$--$\mathit{IDVx}$					\\ 
\hline
\end{tabular}
 
\label{fig:ABCD_CRboth}
\end{table}
 
The sources of background that contribute primarily to ID vertices with  $n_{\mathrm{trk}} = 2$ are not exactly the same as those that contribute to higher $n_{\mathrm{trk}}$ ID vertices, thus a second set of validation regions is studied.
The signal contamination is non-negligible for ID vertices with $n_{\mathrm{trk}} > 2$ in events that pass the Muon RoI Cluster trigger and also contain an MSVx. In order to study higher $n_{\mathrm{trk}}$ vertices, the requirement of the MSVx is removed ($\mathit{Trig}$ and $\mathit{Trig},3\mathit{\mbox{-}trk}$ regions) in the second set of validation regions.
In order to further reduce the signal contamination for ID vertices with higher $n_{\mathrm{trk}}$, the requirement on $m_{\mathrm{IDVx}}$ is modified to select an $m_{\mathrm{IDVx}}$ range adjacent to that used for the final signal selection.
The $\mathit{Bkg},3\mathit{\mbox{-}trk}$ and $\mathit{Trig},3\mathit{\mbox{-}trk}$ regions contain background events and events passing the Muon RoI Cluster, respectively, that have ID vertices which pass the full signal selection with the exceptions that the $m_{\mathrm{IDVx}}$ selection is changed to  $1~\GeV < m_{\mathrm{IDVx}} < 3~\GeV$, and the $n_{\mathrm{trk}}$ selection is changed to $n_{\mathrm{trk}}= 3$.
 
The predicted and observed numbers of events in the $\mathit{Val},2\mathit{\mbox{-}trk}$ and $\mathit{Trig},3\mathit{\mbox{-}trk}$ regions are listed in Table~\ref{table:backgroundNumbers}.
The predicted and observed numbers of events in the $\mathit{Val},2\mathit{\mbox{-}trk}$ regions agree within 2\%, and the kinematic distributions for the two-track vertices in the $\mathit{Val},2\mathit{\mbox{-}trk}$ and $\mathit{Bkg},2\mathit{\mbox{-}trk}$ regions are also found to be in good agreement.
The  predicted and observed numbers of events in the $\mathit{Trig},3\mathit{\mbox{-}trk}$ regions agree within 25\%.
 
The event selection used to populate the different regions impacts the jet multiplicity in each region.
The probability to find an IDVx from background is highly correlated with the jet activity in the events, and the effect is more pronounced for higher $n_{\mathrm{trk}}$ ID vertices than for two-track ID vertices.
It is found that scaling to adjust for the differences in the jet multiplicities reduces the difference in the predicted and observed numbers of events in the
$\mathit{Trig},3\mathit{\mbox{-}trk}$ region.
However, the impact of applying the same scaling to the predicted number of background events in the final signal region changed the final background estimate by less than the statistical uncertainty, so this scaling was not ultimately applied.

The largest difference in all background systematic uncertainty studies is observed to be within 25\%, so this value is taken to be the systematic uncertainty of the background estimation.
The sensitivity of the results to this choice is tested by doubling the systematic uncertainty on the background; the resulting change in the predicted and observed limits is negligible, as the number of predicted background events is small.
With a 25\% systematic uncertainty, the predicted number of background events passing the final signal selection is $1.16 \pm 0.18 \, (\mathrm{stat.}) \pm 0.29 \, (\mathrm{syst.})$.
 
\begin{table}[!htb]
\centering
\caption{Numbers used in the background estimation method and for the validation of the estimate, for the observed ($n_{\mathrm{obs}}$) number of events in the $\mathit{Bkg}$, $\mathit{Bkg}$+$\mathit{IDVx}$, and $\mathit{Sig}$--$\mathit{IDVx}$ regions, and the predicted ($n_{\mathrm{pred}}$) and observed numbers of events in the $\mathit{Val},2\mathit{\mbox{-}trk}$, $\mathit{Trig},3\mathit{\mbox{-}trk}$, and $\mathit{Sig}$ regions.  }
\begin{tabular}{ l  c  c   }
\hline
&	\multicolumn{2}{c}{$n_{\mathrm{obs}}$ }						\\
\hline
Region $\mathit{Bkg}$				&	\multicolumn{2}{c}{6,099,660}								\\
Region $\mathit{Bkg}$+$\mathit{IDVx}$			&	\multicolumn{2}{c}{45}									\\
Region $\mathit{Sig}$--$\mathit{IDVx}$			&	\multicolumn{2}{c}{156,805}								\\
\hline
&	$n_{\mathrm{pred}}$									& $n_{\mathrm{obs}}$	\\
\hline
Region $\mathit{Val},2\mathit{\mbox{-}trk}$		&	$11,269\pm 46 \, \mathrm{(stat.)}	$					& 11,470				\\
Region $\mathit{Trig},3\mathit{\mbox{-}trk}$	&	$1750 \pm 64 \, (\mathrm{stat.})$						& 2132				\\
Region $\mathit{Sig}$				&	$1.16 \pm 0.18 \, (\mathrm{stat.}) \pm 0.29 \, (\mathrm{syst.})$	&	1				\\
 
\hline
 
\end{tabular}
\label{table:backgroundNumbers}
\end{table}
 
 
\section{Systematic uncertainties in signal predictions}
\label{sec:syst}
 
Several sources of systematic uncertainty in the signal selection efficiency are considered.
The dominant uncertainty is due to the difference in performance of the LRT and ID displaced vertex reconstruction algorithms between data and MC simulation.
 
\subsection{Systematic uncertainties in large-radius tracking and displaced vertex reconstruction}
 
To assess the systematic uncertainty of the ID vertex reconstruction efficiency due to the modeling of the large-radius tracking and vertex reconstruction, the rates of displaced vertices consistent with $K^0_{\mathrm{S}} \rightarrow \pi^+ \pi^-$ decays are compared between data and multijet simulation.
The uncertainty is estimated by examining the variations between data and simulation in the $K^0_{\mathrm{S}}$ yield as function of vertex radius.
 
Events in data and MC samples are selected by requiring the presence of a PV.
The events used in data are in the selected subset of events that underwent the reconstruction of displaced tracks and vertices.
The same reconstruction is performed on the multijet MC sample, and the simulated events are reweighted to correctly reproduce the distribution of the mean number of interactions per bunch crossing in the data.
To minimize the statistical uncertainty, no additional event-level requirements are applied.
From the selected events, candidate $K^0_{\mathrm{S}}$ vertices are identified by requiring that the vertices have a decay length greater than 15~mm, have exactly two tracks, and have an invariant mass in the region 450 to 550 MeV.
After the last two selection criteria are applied, the vast majority of selected vertices originate from $K^0_{\mathrm{S}}$ decays and the possibility of signal contamination is minimized.
The kinematic distributions of candidate $K^0_S$ vertices are compared between data and MC and are found to have good agreement within statistical uncertainties.
 
The number of $K^0_{\mathrm{S}}$ candidate vertices found in data and simulation are binned by their decay radius $R$.
To achieve a better estimate of the number of $K^0_{\mathrm{S}}$ in each bin the background contribution is computed from the sidebands of the invariant mass distribution, from 350 to 450 MeV and from 550 to 650 MeV, and subtracted from number of $K^0_{\mathrm{S}}$ candidates in the region 450 to 550 MeV.
 
Tracks originating from a $K^0_{\mathrm{S}}$ decay can be reconstructed by either the standard tracking or the LRT algorithm.
The data are normalized such that the number of $K^0_{\mathrm{S}}$ vertices with two standard tracks is the same between data and simulation.
This accounts for any differences that may exist between data and simulation in the total number of $K^0_{\mathrm{S}}$ decays.
The vertex yields of $K^0_{\mathrm{S}}$ with two large-radius tracks are compared between data and MC simulation, and the largest difference in the ratio of data to MC is found to be $20\%$.
For vertices in the signal region, the effect of this tracking inefficiency is reduced due to the high multiplicity of tracks present in the vertices.
The uncertainty in the vertex reconstruction efficiency is taken to be $20\%$ and is applied as an uncertainty of the global selection efficiency.
 
\subsection{Other systematic uncertainties}
The uncertainty of the integrated luminosity measurement is 2.2\% \cite{ATLAS-CONF-2019-021_fix}, obtained using the LUCID-2 detector \cite{Avoni_2018} for the primary luminosity measurements.
 
The systematic uncertainties of the Muon RoI Cluster trigger efficiency and the MSVx reconstruction efficiency are examined in Ref.~\cite{EXOT-2017-05}.
The systematic uncertainty due to the trigger scale factors is evaluated by varying the scale factors up and down by the uncertainty of the scale factor fit, and comparing the trigger efficiency resulting from the modified scale factors with the trigger efficiency from the nominal scale factor.
The systematic uncertainties are evaluated separately in the barrel and the endcaps.
Similar methods are used to determine the impact of the pileup uncertainty and the systematic uncertainty from the PDF used to generate the signal MC events on the trigger efficiency.
The largest relative uncertainty in the trigger efficiency in any given mass point used in this analysis is found to be 4.8\%.
 
These methods are also used to evaluate the impact of the pileup and PDF uncertainties on the MSVx reconstruction efficiency, and the largest relative uncertainty for any mass point in the barrel or endcap is found to be 5.5\%.
 
 
\section{Results}
\label{sec:result}
 
 
One event is observed that passes the full signal selection, consistent with the estimated background of $1.16 \pm 0.18 \text{ (stat.)}\pm 0.29 \text{ (syst.)}$ events.
 
Upper limits at the 95\% confidence level (CL) are set on the production cross section times branching ratio, for various signal mass hypotheses, following the $\text{CL}_\text{S}$ prescription~\cite{ALRead} with a profile likelihood ratio used as the test statistic.
An asymptotic approach~\cite{Cowan:2010js} is used to compute the $\text{CL}_\text{S}$ value.
This method was tested and found to give results consistent with those obtained from ensemble tests.
A Poisson probability term describing the total number of observed events is used, and the systematic uncertainties of the signal efficiency, background estimation, and luminosity are treated as nuisance parameters and are assigned Gaussian constraints.
 
To evaluate the efficiency of the full event selection as a function of the proper lifetime of the LLP, a reweighting procedure is used, following the method described in Ref.~\cite{CalRatio2016}.
The extrapolated efficiency from the 5~m mean lab-frame decay length sample agrees with the efficiency from the 9~m mean lab-frame decay length sample within the combined statistical uncertainty from the efficiency and extrapolation computations.
 
\begin{figure}[!htb]
\centering
\subfloat[]{\label{subfig:limits125}\includegraphics[width=0.495\textwidth]{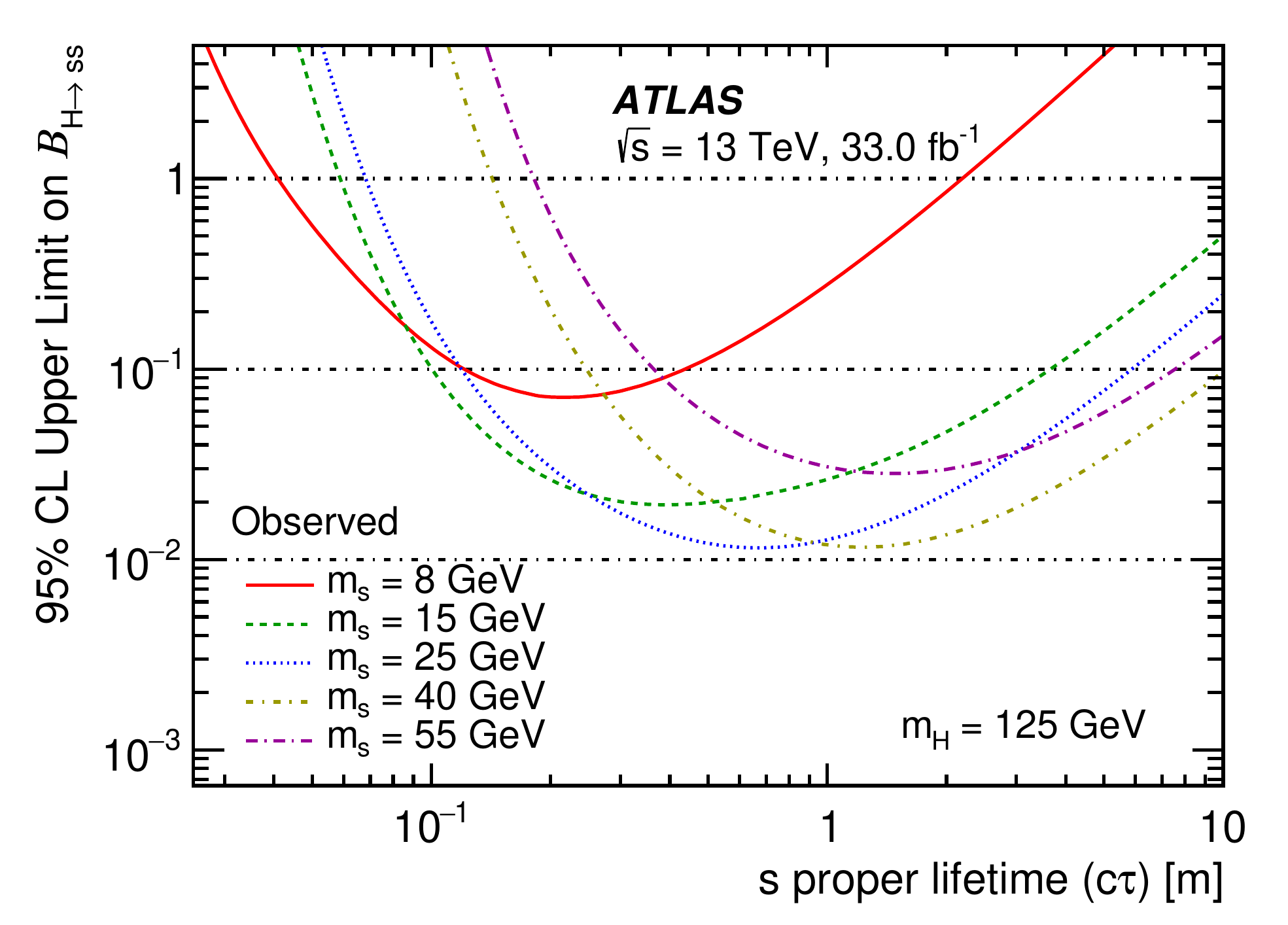}}
\\
\subfloat[]{\label{subfig:limits200_400}\includegraphics[width=0.495\textwidth]{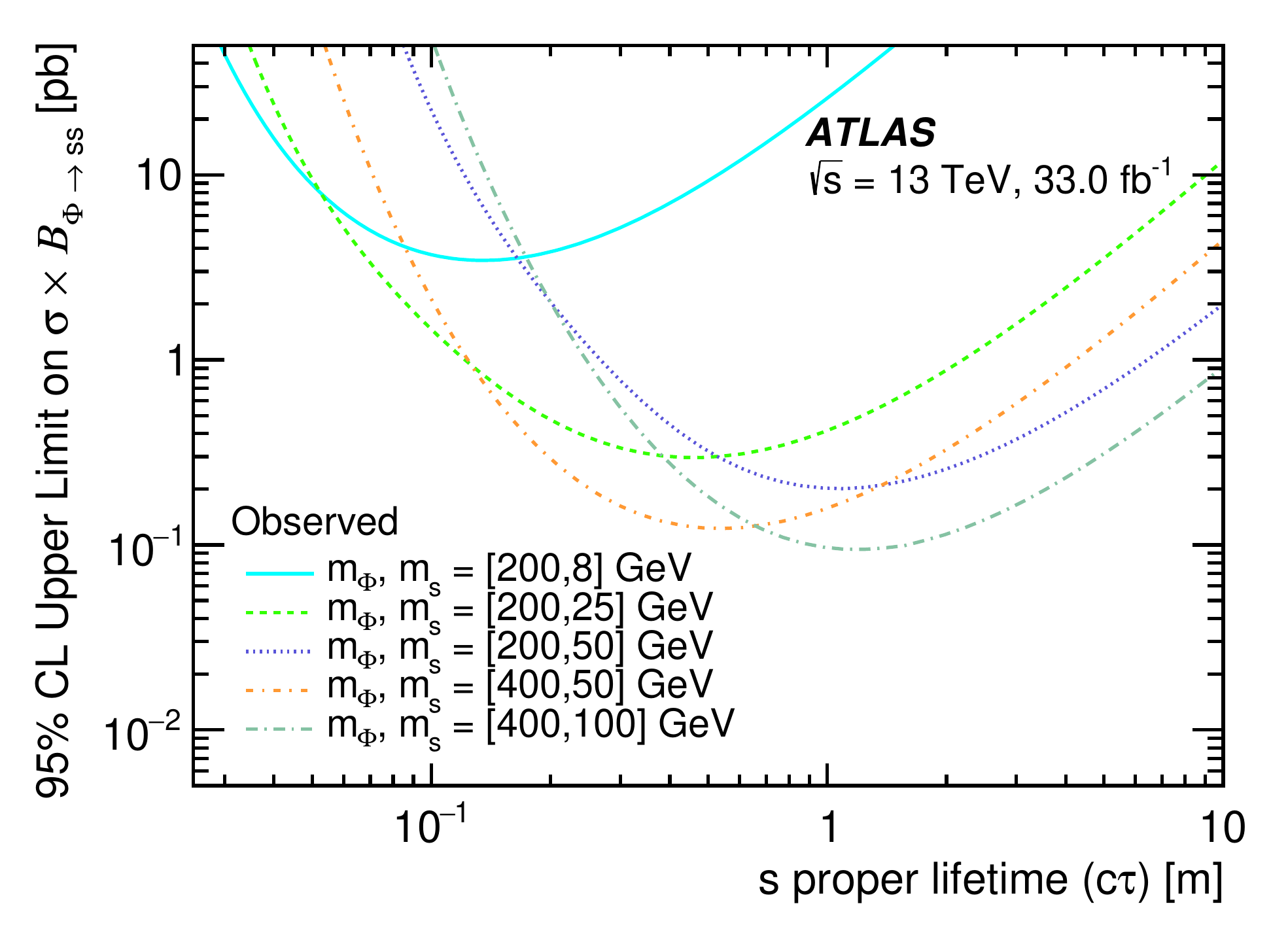}}
\,
\subfloat[]{\label{subfig:limits600_1000}\includegraphics[width=0.495\textwidth]{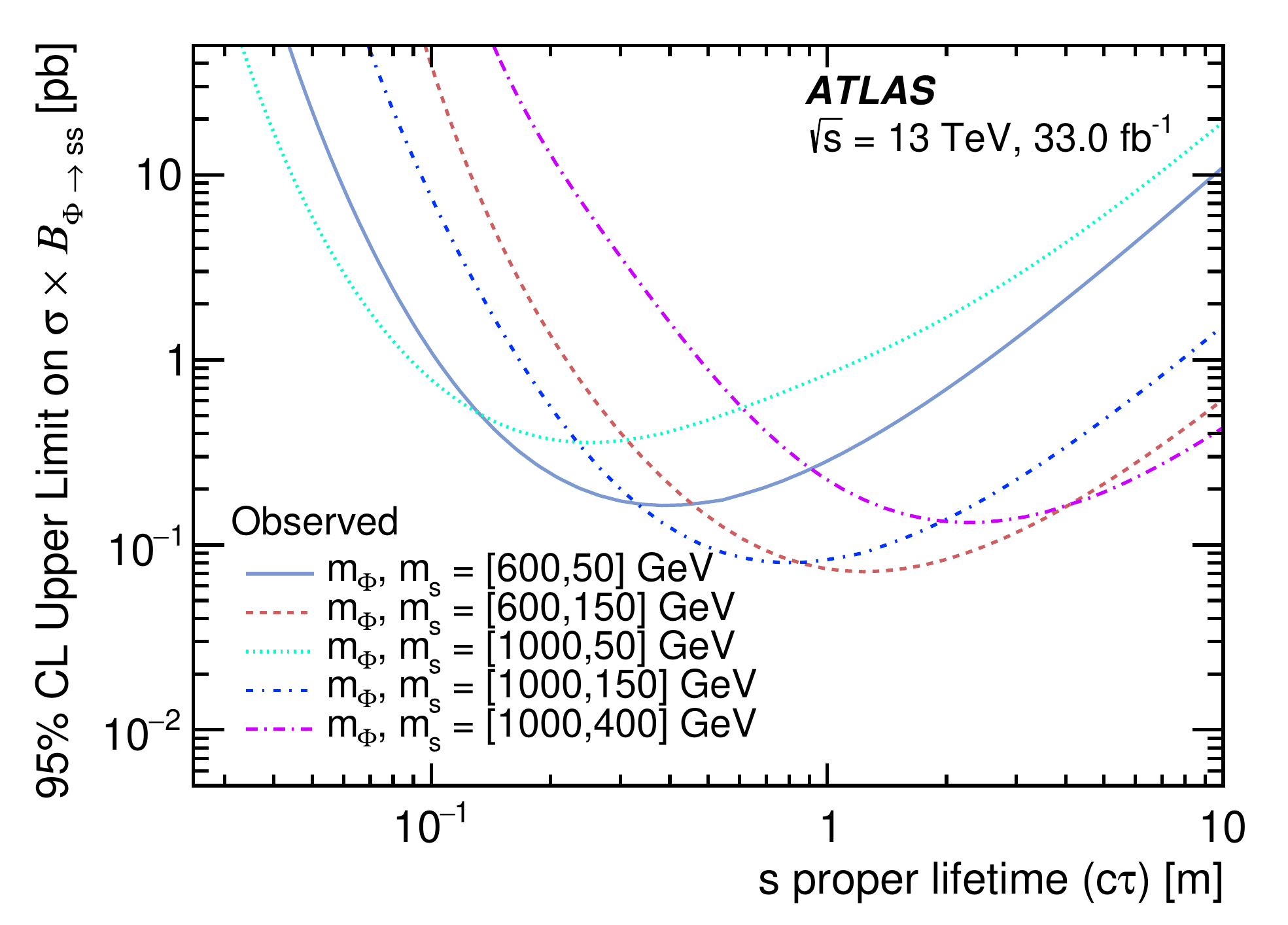}}
\caption{$\text{CL}_{\text{S}}$ upper limits at 95\% CL on the branching ratio $B_{H \to ss}$ for (a) $m_H = 125$~GeV, and on the cross section times branching ratio $\sigma \times B_{\phi \to ss}$ for (b) $m_{\Phi} = 200$ -- $400$~GeV, and (c) $m_{\Phi} = 600$ -- $1000$~GeV.}
\label{fig:Limits}
\end{figure}

The observed limits for all benchmark models considered are summarized in Figure~\ref{fig:Limits}.
For the $m_H = 125$~GeV mediator, the SM Higgs boson gluon--gluon fusion production cross section of 49~pb~\cite{deFlorian:2016spz} at $13$~TeV is assumed and limits on the branching ratio $B_{H \to ss}$ are shown.
The observed limits are consistent with the expected limits within $\pm 1\sigma$ and significantly extend the limits set by previous displaced jet searches~\cite{CalRatio2016,EXOT-2017-05} for low scalar masses and short lifetimes.
The analysis is most sensitive to $c\tau$ values between 0.1~m and 3~m, where equal numbers of LLP decays are found in the fiducial volumes of the ID and the MS.
 
\subsection{Combination of results with other displaced jets searches}
 
The results presented in this paper are complementary to the CR~\cite{CalRatio2016} and MS~\cite{EXOT-2017-05} analyses, which set limits on the same benchmark models as used in this analysis.
The results derived above are combined with the results from the CR and MS analyses to provide increased sensitivity over a greater range of proper lifetimes.
 
The orthogonality of the search presented in this paper (the ID analysis) and the CR analysis is ensured by vetoing on the hadronic calorimeter triggers in both the data and signal MC events as described in Section~\ref{sec:select}.
The MS analysis is separated into the 1-MSVx plus missing transverse momentum (MS1) and 2-MSVx (MS2) channels.  
To ensure orthogonality between the ID and MS analyses, only the 2-MSVx channel is used in the combination.
Across all MC signal samples in this channel, only a few events are found which contain both an IDVx and two MS vertices passing all selections used by the respective analyses, and these events are explicitly vetoed for the combination.
There are no events found in data which pass the full selection of both the ID and MS2 analyses.
The orthogonality between the CR and MS analyses has been verified in Ref.~\cite{CalRatio2016}.
This ensures that the final selected MC signal events and data events in all three analyses used in the combination are statistically independent.
 
The combination is performed using a simultaneous fit of the likelihood functions of each analysis.
The signal strength is correlated between all three likelihoods, as is the nuisance parameter for the luminosity uncertainties.
The signal uncertainties are treated as uncorrelated, since they are dominated by different experimental uncertainties in each search.
The background estimates in each analysis are derived using independent data-driven methods and are therefore not correlated.
 
As in the individual searches, the asymptotic approach is used to compute the $\text{CL}_{\text{S}}$ value, and the limits are defined by the region excluded at 95\% CL.
The limits are calculated using a global fit, where the overall likelihood function is the product of the individual likelihood functions of the searches to be combined.
The limits are calculated separately at each $c\tau$ point, and at each point the signal efficiency is scaled by the result of the lifetime extrapolation procedure.
 
\begin{figure}[!htb]
\centering
\subfloat[]{\label{subfig:limit125_15}\includegraphics[width=0.495\textwidth]{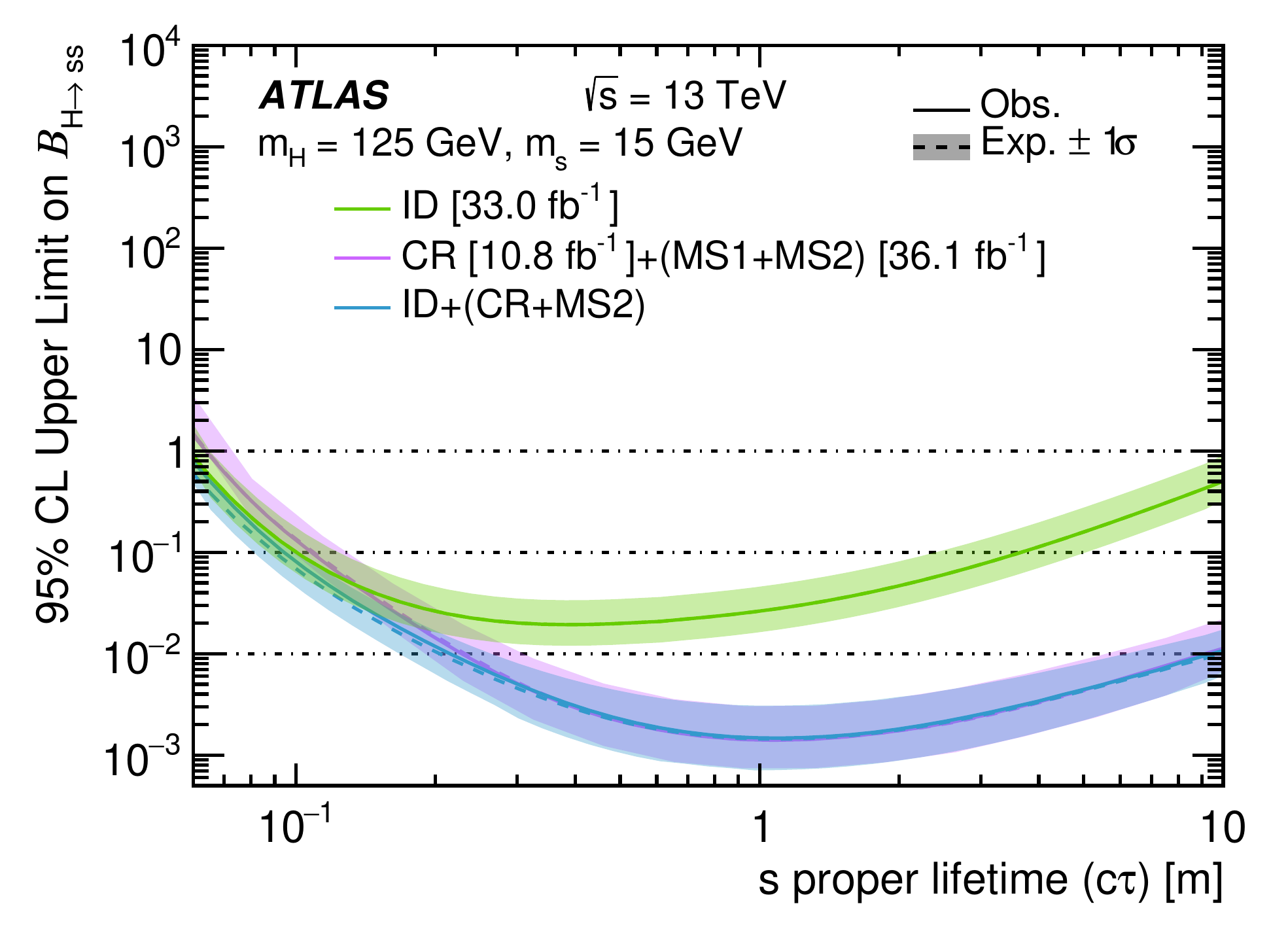}}
\,
\subfloat[]{\label{subfig:limit125_25}\includegraphics[width=0.495\textwidth]{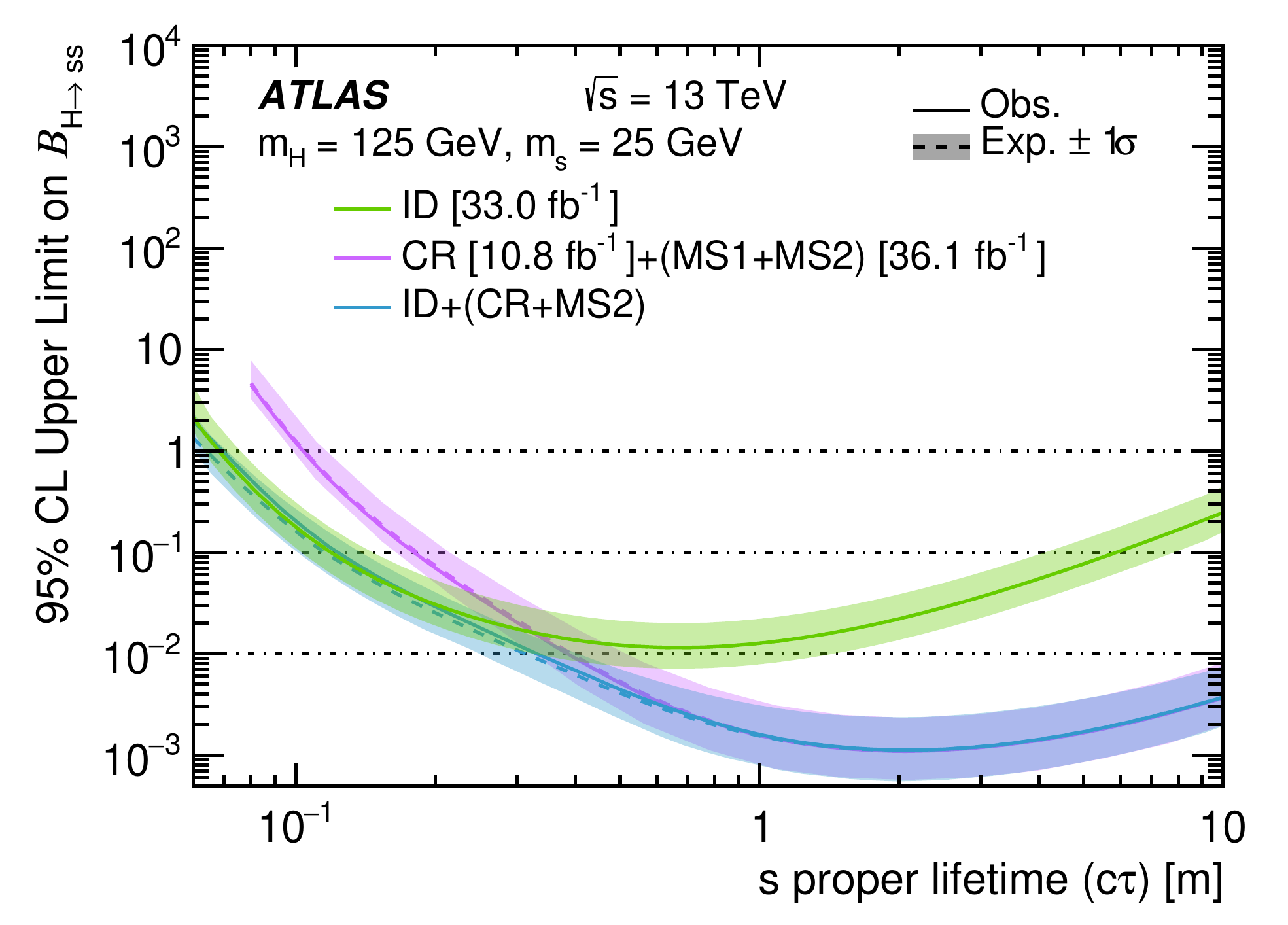}}
\\
\subfloat[]{\label{subfig:limit125_40}\includegraphics[width=0.495\textwidth]{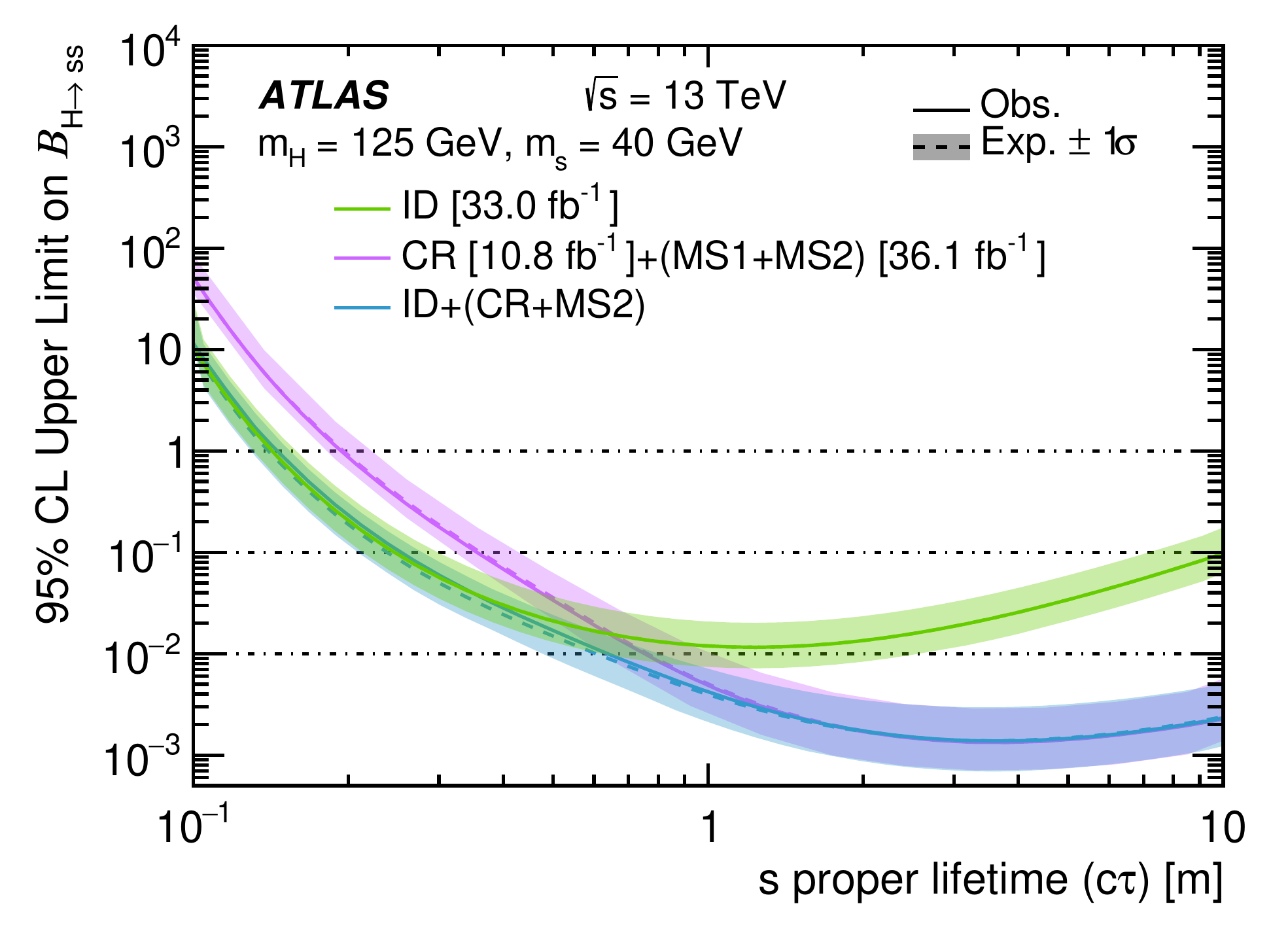}}
\,
\subfloat[]{\label{subfig:limit125_55}\includegraphics[width=0.495\textwidth]{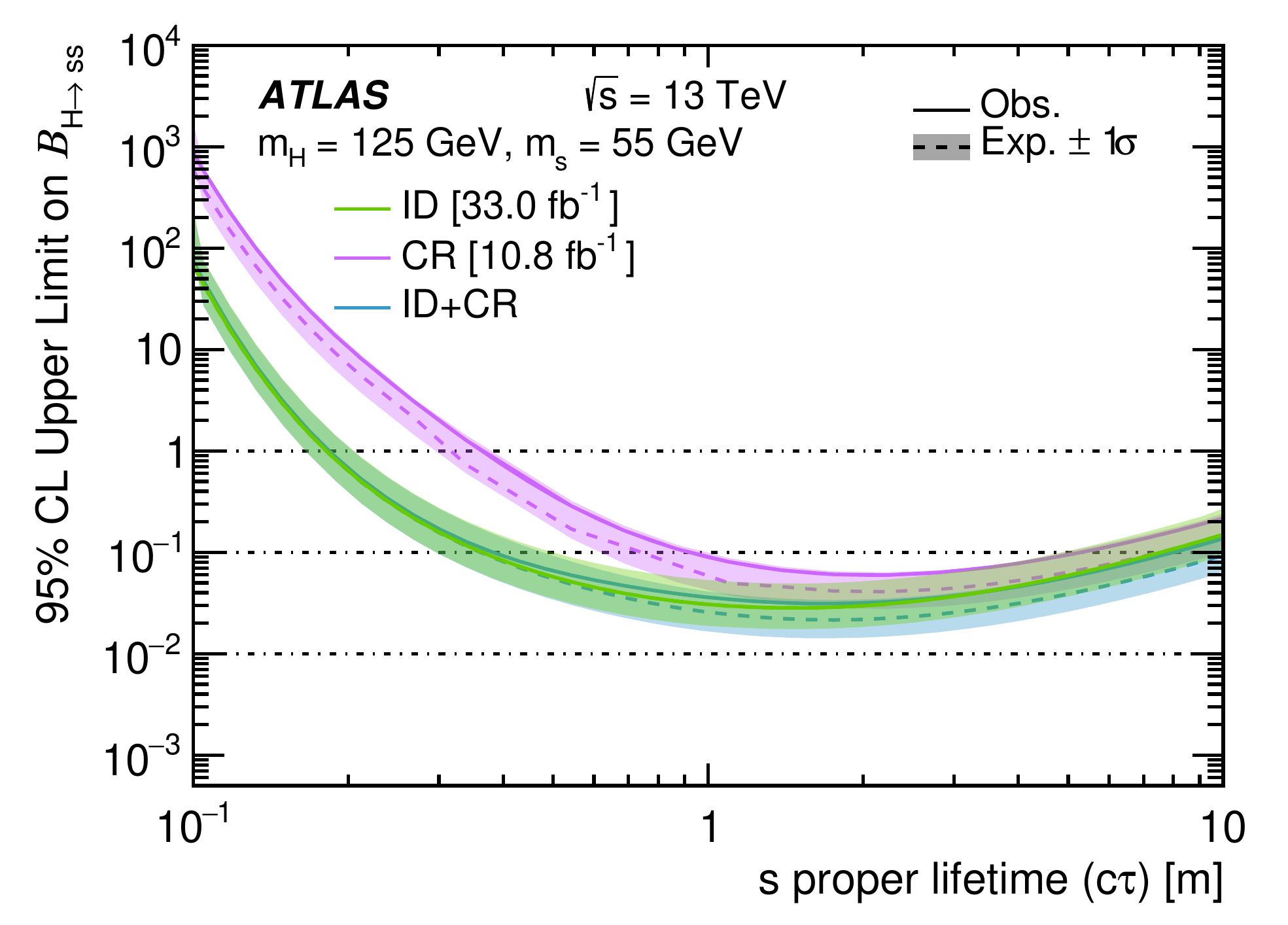}}
\caption{Combined limits from this analysis (ID) and the CR and MS analyses for $m_H = 125$~GeV decaying into (a) 15, (b) 25, (c) 40, and (d) 55~\GeV~mass scalars.
The expected limit is shown as a dashed line with shading for the $\pm 1\sigma$ error band and the observed limit is shown with a solid line.
The combination of the CR and MS analyses used both the MS1 and MS2 channels for $m_{H} = 125$~GeV, but due to orthogonality considerations only the MS2 channel was used when performing the combination with the ID analysis.
The MS analysis did not place limits on the 55~\GeV~LLP mass point (d) so the combined limits use the results of the ID and CR analyses only.
}
\label{fig:LimitCompare_mH125}
\end{figure}

\begin{figure}[!htb]
\centering
\subfloat[]{\label{subfig:limit200_25}\includegraphics[width=0.495\textwidth]{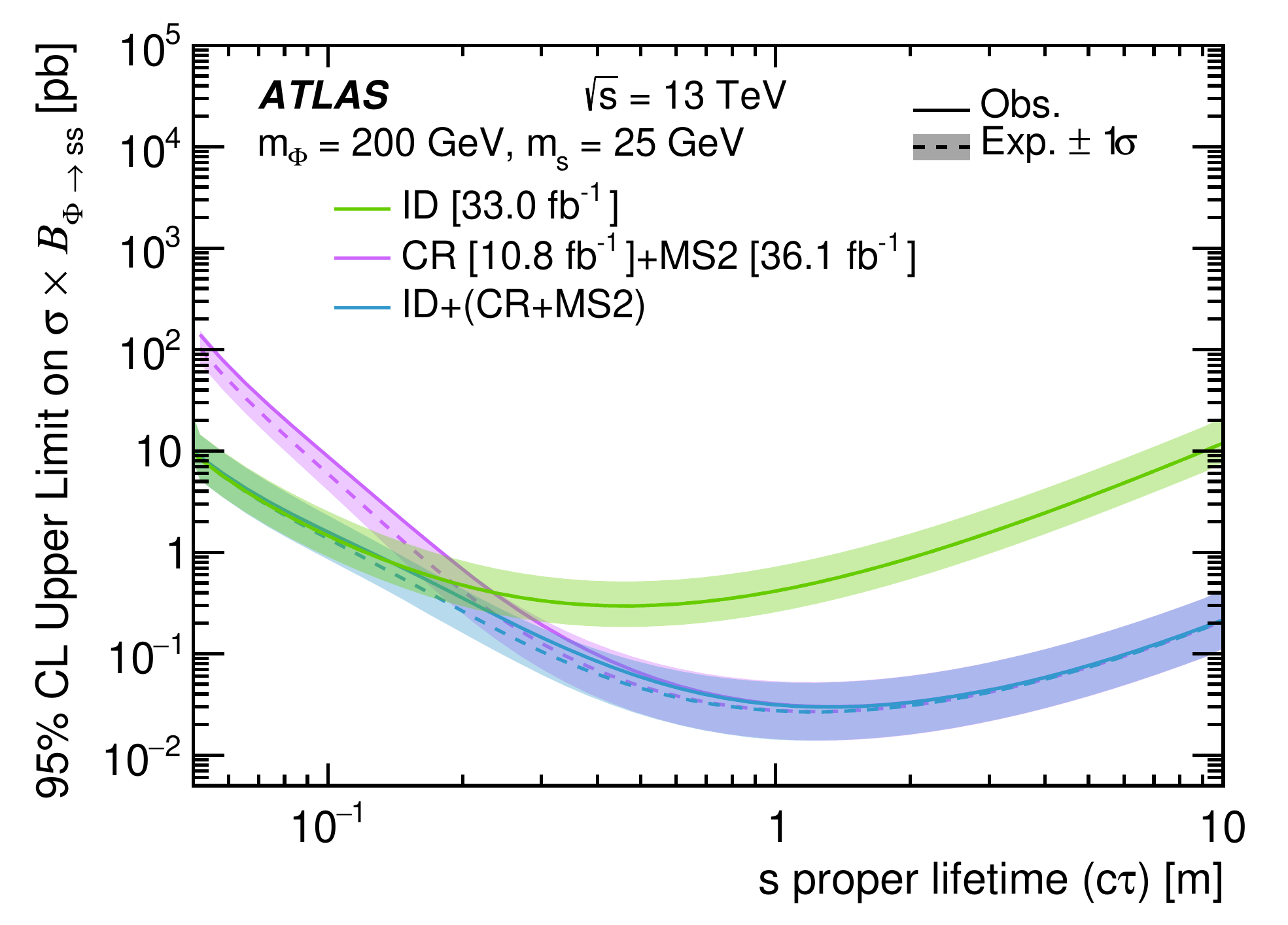}}
\,
\subfloat[]{\label{subfig:limit200_50}\includegraphics[width=0.495\textwidth]{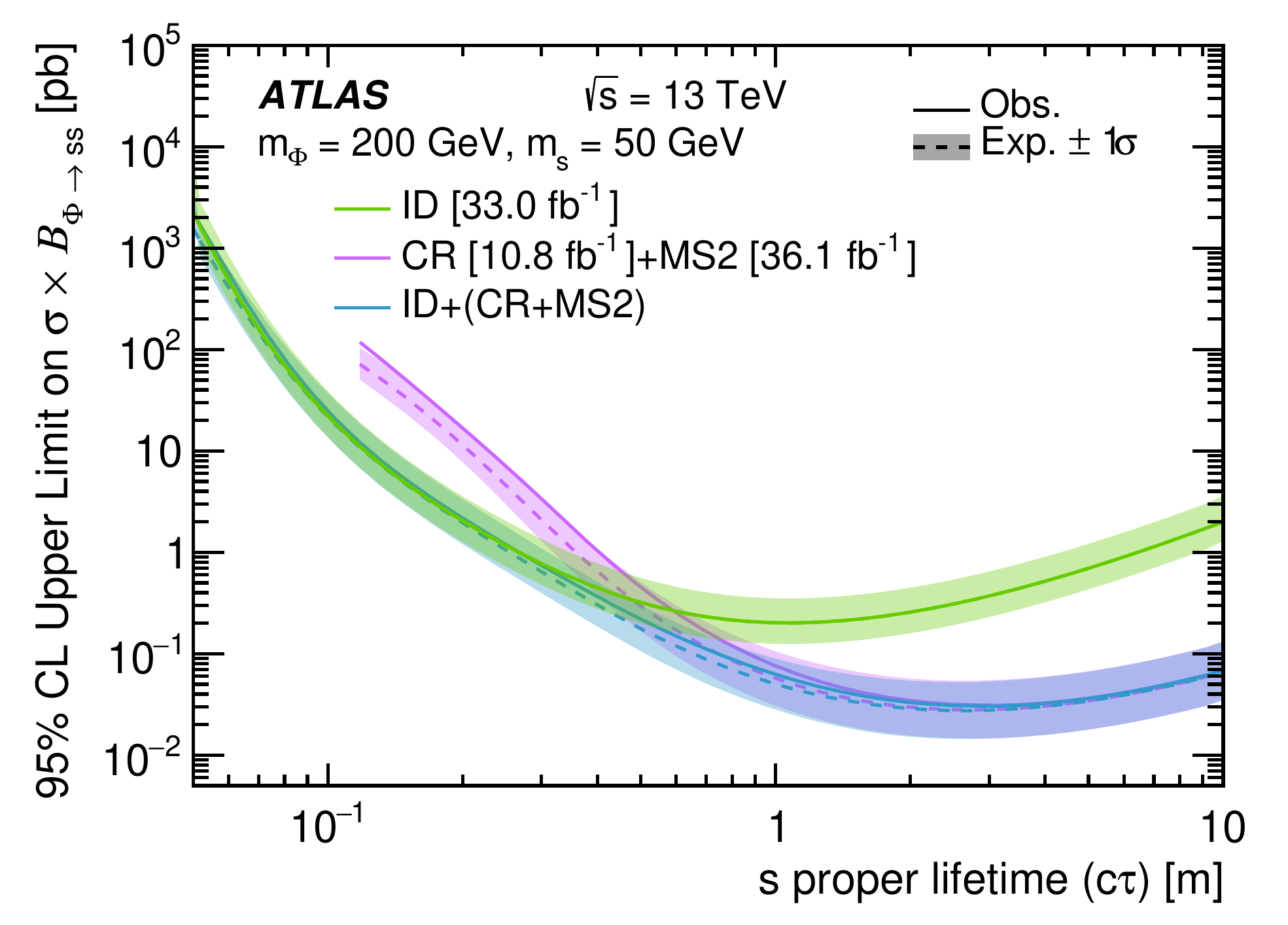}}
\caption{Combined limits from this analysis (ID) and the CR and MS analyses for benchmark models with $m_{\Phi}=200$~GeV decaying into (a) 25 and (b) 50~\GeV~mass scalars.
The expected limit is shown as a dashed line with shading for the $\pm 1\sigma$ error band and the observed limit is shown with a solid line.
}
\label{fig:LimitCompare_mH200}
\end{figure}

Figures~\ref{fig:LimitCompare_mH125} and~\ref{fig:LimitCompare_mH200} show the observed and expected limits for the ID analysis, as well as the combination of the CR and MS analyses both with and without their combination with the ID analysis.
For the models with $m_{H} = 125$~GeV or $m_{\Phi} = 200$~GeV and 8~\GeV~$ < m_s \leq 55$~GeV, the ID analysis has greater sensitivity than the combination of the CR and MS analyses for proper lifetimes ranging from 0.05~m up to 0.7~m.
Although the IDVx reconstruction efficiency diminishes with decreasing LLP masses, requiring both a good MSVx and a good IDVx suppresses the background in the final signal region and allows stronger limits to be set at low $c\tau$ for these mass points.
 
Table~\ref{table:Excl} summarizes the proper lifetime ranges excluded by the combinations for the $m_H = 125$~GeV mediator, assuming a 10\% branching ratio for $H \rightarrow ss$ and using the SM Higgs boson gluon--gluon fusion production cross section.
 
\begin{table}[htb!]
\centering
\caption{Ranges of proper lifetimes excluded at 95\% CL for the $m_H = 125$~GeV benchmark model assuming a 10\% branching ratio for $H \rightarrow ss$. The $m_s = 55$~GeV exclusion range uses the results of the ID and CR analyses only.}
\begin{tabular}{ r | l  p{2cm}  p{4cm} }
\hline
$m_s$~[GeV]	  & Excluded $c\tau$ range at 95\% CL~[m]	\\
\hline
8~~~~~~~             & ~~~~~~~~~~~~~~~~~~~~~0.06--15 \\
15~~~~~~~            & ~~~~~~~~~~~~~~~~~~~~~0.09--64  \\
25~~~~~~~            & ~~~~~~~~~~~~~~~~~~~~~0.12--116  \\
40~~~~~~~            & ~~~~~~~~~~~~~~~~~~~~~0.26--197  \\
55~~~~~~~            & ~~~~~~~~~~~~~~~~~~~~~0.39--8.1  \\
\hline
\end{tabular}
\label{table:Excl}
\end{table}
 
The combination of the results was also explored for signal points with $m_{\Phi} \geq 400$~GeV.
The sensitivity of this analysis increases as the $\Phi$- and $s$-boson masses increase, due to the increased fraction of reconstructed ID vertices passing the IDVx selection requirements, as shown in Figure~\ref{fig:SvB_massnTrks}.
However, the gain in sensitivity with increasing masses is much larger for the CR-analysis.
Thus, the addition of the results from this analysis do not noticeably extend the combined exclusion limits for signal points with $m_{\Phi} \geq 400$~GeV, and they are therefore not shown in this paper.

 
\FloatBarrier

\section{Conclusion}
\label{sec:conclusion}
 
This paper presents a search for pairs of long-lived particles decaying in the ATLAS inner tracking detector and muon spectrometer, using 33.0 $\mathrm{fb}^{-1}$ of $\sqrt{s} = 13$~TeV proton--proton collision data which were collected at the LHC in 2016.
Benchmark HS models are studied, using a scalar mediator that ranges in mass from 125 to 1000 \GeV, decaying into pairs of long-lived scalars ranging in mass from 8 to 400 \GeV, depending on the mass of the mediator.
The search presented focuses on the topology consisting of one displaced hadronic decay in the inner detector and one in the muon spectrometer.
The search employs dedicated techniques to reconstruct both the displaced inner detector and muon spectrometer hadronic vertices.
A data-driven background estimation is performed, which predicts approximately one event in the signal region from background sources.
One event is found in the signal region, and limits are set on the various signal mass points.
This search has a greater sensitivity for low-mass, long-lived scalars at shorter lifetimes than previously published searches.
The limits resulting from the combination of this search with the previous ATLAS searches for long-lived particles are the most stringent thus far on the branching ratios from the Higgs boson to several low-mass scalars, and on the cross section times branching ratio for a 200 \GeV~mass $\Phi$ decaying into long-lived scalars with masses of 25 and 50~\GeV.

\section*{Acknowledgments}
 

We thank CERN for the very successful operation of the LHC, as well as the
support staff from our institutions without whom ATLAS could not be
operated efficiently.
 
We acknowledge the support of ANPCyT, Argentina; YerPhI, Armenia; ARC, Australia; BMWFW and FWF, Austria; ANAS, Azerbaijan; SSTC, Belarus; CNPq and FAPESP, Brazil; NSERC, NRC and CFI, Canada; CERN; CONICYT, Chile; CAS, MOST and NSFC, China; COLCIENCIAS, Colombia; MSMT CR, MPO CR and VSC CR, Czech Republic; DNRF and DNSRC, Denmark; IN2P3-CNRS and CEA-DRF/IRFU, France; SRNSFG, Georgia; BMBF, HGF and MPG, Germany; GSRT, Greece; RGC and Hong Kong SAR, China; ISF and Benoziyo Center, Israel; INFN, Italy; MEXT and JSPS, Japan; CNRST, Morocco; NWO, Netherlands; RCN, Norway; MNiSW and NCN, Poland; FCT, Portugal; MNE/IFA, Romania; MES of Russia and NRC KI, Russia Federation; JINR; MESTD, Serbia; MSSR, Slovakia; ARRS and MIZ\v{S}, Slovenia; DST/NRF, South Africa; MINECO, Spain; SRC and Wallenberg Foundation, Sweden; SERI, SNSF and Cantons of Bern and Geneva, Switzerland; MOST, Taiwan; TAEK, Turkey; STFC, United Kingdom; DOE and NSF, United States of America. In addition, individual groups and members have received support from BCKDF, CANARIE, Compute Canada and CRC, Canada; ERC, ERDF, Horizon 2020, Marie Sk{\l}odowska-Curie Actions and COST, European Union; Investissements d'Avenir Labex, Investissements d'Avenir Idex and ANR, France; DFG and AvH Foundation, Germany; Herakleitos, Thales and Aristeia programmes co-financed by EU-ESF and the Greek NSRF, Greece; BSF-NSF and GIF, Israel; CERCA Programme Generalitat de Catalunya and PROMETEO Programme Generalitat Valenciana, Spain; G\"{o}ran Gustafssons Stiftelse, Sweden; The Royal Society and Leverhulme Trust, United Kingdom.
 
The crucial computing support from all WLCG partners is acknowledged gratefully, in particular from CERN, the ATLAS Tier-1 facilities at TRIUMF (Canada), NDGF (Denmark, Norway, Sweden), CC-IN2P3 (France), KIT/GridKA (Germany), INFN-CNAF (Italy), NL-T1 (Netherlands), PIC (Spain), ASGC (Taiwan), RAL (UK) and BNL (USA), the Tier-2 facilities worldwide and large non-WLCG resource providers. Major contributors of computing resources are listed in Ref.~\cite{ATL-GEN-PUB-2016-002}.
 

\printbibliography

\clearpage
\input{atlas_authlist} 
\end{document}

%% file: atlas_authlist.tex
 
\begin{flushleft}
{\Large The ATLAS Collaboration}

\bigskip

G.~Aad$^\textrm{\scriptsize 102}$,    
B.~Abbott$^\textrm{\scriptsize 129}$,    
D.C.~Abbott$^\textrm{\scriptsize 103}$,    
A.~Abed~Abud$^\textrm{\scriptsize 36}$,    
K.~Abeling$^\textrm{\scriptsize 53}$,    
D.K.~Abhayasinghe$^\textrm{\scriptsize 94}$,    
S.H.~Abidi$^\textrm{\scriptsize 167}$,    
O.S.~AbouZeid$^\textrm{\scriptsize 40}$,    
N.L.~Abraham$^\textrm{\scriptsize 156}$,    
H.~Abramowicz$^\textrm{\scriptsize 161}$
% ...
\bigskip
\\

% ...
$^{36}$CERN, Geneva; Switzerland.\\
% ...
$^{40}$Niels Bohr Institute, University of Copenhagen, Copenhagen; Denmark.\\
% ...
$^{53}$II. Physikalisches Institut, Georg-August-Universit\"{a}t G\"ottingen, G\"ottingen; Germany.\\
% ...
$^{94}$Department of Physics, Royal Holloway University of London, Egham; United Kingdom.\\
% ...
$^{102}$CPPM, Aix-Marseille Universit\'e, CNRS/IN2P3, Marseille; France.\\
$^{103}$Department of Physics, University of Massachusetts, Amherst MA; United States of America.\\
% ...
$^{129}$Homer L. Dodge Department of Physics and Astronomy, University of Oklahoma, Norman OK; United States of America.\\
% ...
$^{156}$Department of Physics and Astronomy, University of Sussex, Brighton; United Kingdom.\\
% ...
$^{161}$Raymond and Beverly Sackler School of Physics and Astronomy, Tel Aviv University, Tel Aviv; Israel.\\
% ...
$^{167}$Department of Physics, University of Toronto, Toronto ON; Canada.\\
% ...

\end{flushleft}
